\pdfoutput=1

\documentclass[letterpaper, 10 pt, conference]{ieeeconf}

\usepackage{graphicx}
\usepackage{booktabs}
\usepackage{adjustbox}
\usepackage{wrapfig}
\usepackage{lipsum}
\usepackage{amssymb}
\usepackage{amsmath}
\usepackage{siunitx}
\usepackage{mathtools}
\usepackage{xcolor}
\usepackage{hyperref}
\let\labelindent\relax
\usepackage[inline]{enumitem}
\usepackage[ruled,vlined,linesnumbered]{algorithm2e}
\usepackage[noend]{algpseudocode}
\usepackage{ifthen}
\usepackage{dsfont}
\usepackage[backend=biber,style=numeric-comp,sorting=none,maxbibnames=99]{biblatex}
\usepackage{balance}
\usepackage{dirtytalk}
\usepackage{afterpage}

\usepackage{times}

\allowdisplaybreaks

\newcommand{\E}{\mathbb{E}}

\newcommand{\N}{\mathbb{N}}

\newcommand{\Prob}{\mathbb{P}}

\newcommand{\ra}{\rightarrow}
\newcommand{\R}{\mathbb{R}}

\newtheorem{remark}{Remark}
\newtheorem{assumption}{Assumption}
\newtheorem{proposition}{Proposition}
\newtheorem{definition}{Definition}
\newtheorem{lemma}{Lemma}
\newtheorem{theorem}{Theorem}

\newcommand{\graphMod}{G}

\newcommand{\nodesMod}{I}

\newcommand{\arcsMod}{A}

\newcommand{\arcsLoadConstraintSet}{\mathcal{W}}
\newcommand{\arcLoadMod}{w}

\newcommand{\latency}{s}
\newcommand{\toll}{p}

\newcommand{\costToGo}{z}

\newcommand{\routeSetMod}{\textbf{R}}
\newcommand{\routes}{\routeSetMod}

\newcommand{\depth}{\ell}

\newcommand{\height}{m}

\newcommand{\latencyTotal}{L}

\newcommand{\nodeLoadIn}{g}

\usepackage{todonotes}
\usepackage{bbm}

\bibliography{references}

\IEEEoverridecommandlockouts
\title{\LARGE \bf Parameter Estimation in Optimal Tolling for Traffic Networks Under the Markovian Traffic Equilibrium
}

\author{Chih-Yuan Chiu$^{1}$ and Shankar Sastry$^{1}$
\thanks{Supported by NSF Grant 2031899, Collaborative Research: Transferable, Hierarchical, Expressive, Optimal, Robust, Interpretable Networks.
}
\thanks{$^{1}$Department of Electrical Engineering and Computer Sciences, University of California, Berkeley, CA 94720 (emails: \texttt{\{chihyuan\_chiu, sastry\} at berkeley dot edu}).}%
}

\begin{document}

\maketitle

\thispagestyle{empty}
\pagestyle{empty}




\begin{abstract}
Tolling, or congestion pricing, has emerged as an effective tool for preventing gridlock in traffic systems. However, tolls are currently mostly designed on route-based traffic assignment models (TAM), which may be unrealistic and computationally expensive. Existing approaches also impractically assume that the central tolling authority can access latency function parameters that characterize the time required to traverse each network arc (edge), as well as the entropy parameter $\beta$ that characterizes commuters' stochastic arc-selection decisions on the network. To address these issues, this work formulates an online learning algorithm that simultaneously refines estimates of linear arc latency functions and entropy parameters in an arc-based TAM, while implementing tolls on each arc to induce equilibrium flows that minimize overall congestion on the network. We prove that our algorithm incurs regret upper bounded by $O(\sqrt{T} \ln(T) |\arcsMod| \max\{|\nodesMod| \ln(|\arcsMod|/|\nodesMod|), B \})$, where $T$ denotes the total iteration count, $|\arcsMod|$ and $|\nodesMod|$ denote the total number of arcs and nodes in the network, respectively, and $B$ describes the number of arcs required to construct an estimate of $\beta$ (usually $\ll |I|$). Finally, we present numerical results on simulated traffic networks that validate our theoretical contributions.
\end{abstract}

\section{Introduction}
\label{sec: Introduction}

Modern transportation systems are often plagued with congestion, induced by commuters who select latency-minimizing routes from their source to their destination in a self-interested manner. Tolling mechanisms, which impose additional prices on each arc (edge) in the network, offer a natural solution to this issue. By appropriately augmenting the overall cost of traveling on particularly congested arcs, effectively implemented tolls can reshape commuters' incentives, and motivate them to make arc selections that reduce the overall network congestion.

Although various traffic assignment and tolling mechanisms have been proposed to regulate congestion on transportation networks, the theoretical guarantees of these approaches, if any, are usually predicated upon unrealistic or impractical modeling assumptions. For instance, \cite{Pulyassary2023CapacityAllocationandPricing, Paccagnan2019IncentivizingEfficientUse, Correa2022NetworkPricing} design traffic assignment schemes or tolls using \textit{route-based traffic assignment models} (TAMs) to capture commuters' navigation decisions, i.e., each commuter is assumed to make a single route selection at their origin, and to refrain from deviating from their selected route at intermediate nodes. Likewise, \cite{Gollapudi2023OnlineLearningforTrafficNavigation} presents an online learning algorithm to infer the unknown latency functions of a traffic network, while performing optimal route assignment over the network in the context of a route-based TAM. Unfortunately, route-based TAMs do not capture the behavior of commuters who re-route halfway to their destination, and can be computationally expensive, since the number of routes in a traffic network can grow exponentially with the number of arcs (edges). In contrast, \cite{Kanekoa2021OptimalCongestionTollingUnderMTE, MaheshwariKulkarni2022DynamicTollingforInducingSociallyOptimalTrafficLoads, Chiu2023ArcbasedTrafficAssignment, Chiu2023DynamicTollinginArcBasedTAMs} investigate commuters' decision making and tolling mechanisms in a traffic network over a stochastic \textit{arc-based TAM}, in which commuters sequentially select among outgoing arcs at each intermediate node from source to destination. In particular, an entropy parameter $\beta > 0$ is used to characterize the degree of irrationality with which the traveler population selects arc sequences, due to the incomplete and imperfect information they possess about the latency cost of each arc. However, these approaches unrealistically assume that the central tolling authority possesses perfect knowledge of $\beta$ and the network latency functions.

To address the above shortcomings, this work presents an online learning algorithm in the framework of a stochastic, arc-based traffic assignment model (TAM), to simultaneously learn the latency function and the entropy parameter, while implementing tolls that become increasingly effective at reducing overall congestion in subsequent iterations. At each iteration, we first implement tolls, constructed during the most recent iteration, on each arc in the network. We then collect the resulting equilibrium traffic flow and latency data from each arc, and apply a regularized least-squares method to update our estimates of the latency function parameters, based on the collected data. In turn, the flow data and latency function estimates can then be used to update our estimate of the entropy parameter $\beta$, using the Principle of Optimism in the Face of Uncertainty. Finally, these improved estimates of the latency function and entropy parameters are used to design an improved tolling strategy for the next iteration. 

We define the stage-wise regret of our algorithm at each iteration $t$ to be the difference between the following two quantities: (a) The overall latency in the network induced by equilibrium flows corresponding to the toll implemented at iteration $t$, and (b) The minimum overall latency attainable by the tolling mechanism if it possessed perfect knowledge of the entropy parameter and each arc latency function. The cumulative regret is then computed by summing the stage-wise regret across all iterations. Our algorithm incurs regret of order $O(\sqrt{T} \ln(T)|\arcsMod| \cdot \max\{|\nodesMod| \ln(|\arcsMod|/|\nodesMod|), B\})$, where $T$ denotes the total iteration count, $|\arcsMod|$ and $|\nodesMod|$ denote the number of arcs and nodes in the network, respectively, and $B$ denotes the number of arcs in the network used to construct the estimate of the entropy parameter $\beta$ at each iteration. 

On a technical level, our algorithm utilizes concepts familiar to the bandits community, such as the regularized least-squares method for latency function estimation \cite{Gollapudi2023OnlineLearningforTrafficNavigation, LattimoreSzepesvari2020BanditAlgorithms, AbbasiYadkori2011ImprovedAlgorithmsforLinearStochasticBandits}, and the Principle of Optimism in the Face of Uncertainty for entropy parameter estimation and toll design \cite{LattimoreSzepesvari2020BanditAlgorithms}. However, the problem formulation and proof methodologies considered in this work differ significantly from the above literature. First, 
in our problem setup, the decision maker's actions are tolls, which induce equilibrium flows through a non-convex map; in turn, the regret is defined from the overall network congestion generated by these equilibrium flows. Similarly, the unknown entropy parameter estimated in our work affects the cumulative regret in a complicated, network structure-dependent manner (see Section \ref{sec: Regret Analysis}, Remark \ref{Remark: Regret Upper Bound}). These complex dependencies between the actions, unknown parameters, and regret preclude the direct use of analysis techniques in the bandit literature. Moreover, whereas the decision-maker in \cite{Gollapudi2023OnlineLearningforTrafficNavigation} estimates latency functions in the context of a route-based TAM and implements optimal flow assignments directly, our work estimates both the latency functions and entropy parameter $\beta$ of an underlying arc-based TAM, and implements tolls, which in turn induce an equilibrium flow from which the regret is computed. In particular, to estimate the entropy parameter $\beta$, we use a novel approximation scheme beyond the methods in \cite{Gollapudi2023OnlineLearningforTrafficNavigation}.

Likewise, various methods have investigated the problem of estimating the entropy parameter of softmax models in the context of traffic assignment models or maximum entropy inverse reinforcement learning \cite{oyama2019prism, oyama2022markovian, Reverdy2015ParameterEstimationInSoftmaxDecisionMakingVariables}. However, these approaches usually use heuristic models to approximate the unknown parameter \cite{oyama2019prism, oyama2022markovian}, or assume that the overall objective can be written as a convex function of the entropy parameter \cite{Reverdy2015ParameterEstimationInSoftmaxDecisionMakingVariables}. These assumptions separate the above methods from our work, since our formulation involves cost and equilibrium models that are highly non-convex in the action variables (tolls) and in the unknown entropy parameter $\beta$.

The following sections are structured as follows. Section \ref{sec: Preliminaries} introduces the traffic network studied throughout the remaining sections, as well as the incentive structures faced by the commuters traversing the network. Section \ref{sec: Main Algorithm} presents our online algorithm. An upper bound for the overall regret incurred by this algorithm is given in Section \ref{sec: Regret Analysis}. Finally, Section \ref{sec: Experiments} presents empirical evidence for the theoretical regret bounds on our algorithm, while Section \ref{sec: Conclusion and Future Work} summarizes our work and presents avenues for future research.

\textit{Notation}: Below, for any $n \in \N$, we denote $[n] := \{1, \cdots, n\}$. For any $n \in \N$ and $i \in [n]$, let $e_i$ denote the $i$-th standard unit vector in the Euclidean space $\R^n$. We set $\textbf{1}\{\cdot\}$ to equal $1$ if the input event occurs, and 0 otherwise.


\section{Preliminaries}
\label{sec: Preliminaries}

\subsection{Setup}
\label{subsec: Setup}

Let $\graphMod = (\nodesMod, \arcsMod)$ be a directed acyclic graph that describes a single-origin single-destination traffic network, with $\nodesMod$ and $\arcsMod$ denoting the set of nodes and the set of arcs, respectively. For each arc $a \in \arcsMod$, we denote the start and end nodes of $a$ by $i_a$ and $j_a$, respectively    . For each node $i \in I$, let $\arcsMod_i^-, \arcsMod_i^+ \subset \arcsMod$ denote the set of incoming and outgoing arcs. 
Let $g_o \geq 0$ denote the traffic flow entering the network $\graphMod$ at each iteration.

To traverse the network, commuters sequentially select from outgoing arcs at each intermediate node, from the origin $o$ to the destination $d$. Each arc $a \in \arcsMod$ is associated with a positive, strictly increasing \textit{latency function} $\latency_a: [0, \infty) \ra [0, \infty)$, which captures the time required to travel through arc $a$ due to congestion produced by the traffic load $\arcLoadMod_a \geq 0$, and a \textit{toll} $\toll_a \geq 0$, the monetary value each traveler must pay to access the arc. Throughout the rest of the paper, we adopt a linear latency model, formally stated as follows\footnote{For an extension of our least-squares-based latency function estimation method to higher-degree polynomial latency functions, please see \cite{Gollapudi2023OnlineLearningforTrafficNavigation}.}.

\begin{assumption}[\textbf{Linear Latency Functions}]
For each arc $a \in \arcsMod$, there exists a coefficient $\theta_a \in \R$ such that $s_a(\arcLoadMod_a) = \theta_a \arcLoadMod_a$.
\end{assumption}

The \textit{cost} $c_a: [0, \infty)^3 \ra [0, \infty)$ on each arc is then obtained by summing the travel time and toll:
\begin{align*}
    c_a(\theta_a, \arcLoadMod_a, \toll_a) &= \latency_a(\arcLoadMod_a) + \toll_a \\
    &= \theta_a \arcLoadMod_a + \toll_a,
\end{align*}
while the \textit{perceived cost} $\tilde c_a$ additionally includes a zero-mean stochastic error term $\delta_a \in \R$ that encapsulates variations in commuters' perception of travel time:
\begin{align*}
    \tilde c_a(\theta_a, \arcLoadMod_a, \toll_a) &= \latency_a(\arcLoadMod_a) + \toll_a + \delta_a \\
    &= \theta_a \arcLoadMod_a + \toll_a + \delta_a,
\end{align*}

At every non-destination node $i \in I \backslash \{d\}$, commuters select among outgoing arcs $a \in A_i^+$ by computing their perceived minimum cost-to-go $\{\tilde z_a \in \R: a \in \arcsMod_i^+\}$ on arc $a$:
\begin{align} \label{eq: LTGPerturbed}
    &\tilde{z}_a(\theta, \arcLoadMod, \toll) \\ \nonumber
    := \hspace{0.5mm} &\tilde c_a(\theta, \arcLoadMod_a, \toll_a) + \E_{\delta} \Big[ \min_{a' \in A_{j_a}^+} \tilde{z}_{a'}(\theta, \arcLoadMod, \toll) \Big], \hspace{2mm} j_a \ne d, \\ \nonumber
    &\tilde{z}_a(\theta, \arcLoadMod, \toll) \\
    := \hspace{0.5mm} &\tilde c_a(\theta, \arcLoadMod_a, \toll_a), \hspace{2mm} j_a = d.
\end{align}
In this work, we adopt the \textit{logit Markovian Model} \cite{Akamatsu1997DecompositionOfPathChoiceEntropy,BaillonCominetti2008MarkovianTrafficEquilibrium}, under which the noise terms $\delta_a$ are described by the Gumbel distribution with scale (or, entropy) parameter $\beta > 0$. As a result, the expected cost-to-go $z_a$ for each arc $a \in \arcsMod$ admits the following closed-form expression:
{\small
\begin{align} \label{Eqn: CostToGo}
    z_a(\theta, \arcLoadMod, \toll) &= c_a (\theta_a, \arcLoadMod_a, \toll_a) - \frac{1}{\beta} \ln\Bigg( \sum_{a' \in \arcsMod_{j_a}^+} e^{-\beta z_{a'}(\theta, \arcLoadMod, \toll)} \Bigg).
\end{align}
}

The corresponding equilibrium flow, called the \textit{Markovian Traffic Equilibrium (MTE) $\bar \arcLoadMod^{\theta, \beta}(\toll) \in \R^{|\arcsMod|}$ corresponding to the latency function parameters $\theta \in \R^{|\arcsMod|}$, entropy parameter $\beta > 0$, and toll vector $\toll \in \R^{|\arcsMod|}$}, is the unique flow vector satisfying the following fixed point equation---For each non-destination node $i \in \nodesMod \backslash \{d\}$ and outgoing arc $a \in \arcsMod_i^+$:
\begin{align*}
    \bar \arcLoadMod_a^{\theta, \beta}(\toll) &= \left(g_i + \sum_{a' \in \arcsMod_{i}^+} \bar \arcLoadMod_{a'}^{\theta, \beta} (\toll) \right) \\
    &\hspace{7.5mm} \cdot  \frac{\exp(-\beta z_a(\theta, \bar \arcLoadMod^{\theta, \beta}(\toll), \toll))}{\sum_{a' \in \arcsMod_i^+} \exp(-\beta z_{a'}(\theta, \bar \arcLoadMod^{\theta, \beta}(\toll), \toll))}, \\
    \bar \arcLoadMod_a^{\theta, \beta}(\toll) &\in \arcsLoadConstraintSet,
\end{align*}
where $g_i := g_o$ if $i = o$ and $g_i = 0$ otherwise, and $\arcsLoadConstraintSet$ is defined as the constraint set that enforces the conservation of traffic flow:
\begin{align} \label{Eqn: Def, W}
    \arcsLoadConstraintSet := &\Bigg\{ \arcLoadMod \in \R^{|\arcsMod|}: \sum_{a \in \arcsMod_i^+} \arcLoadMod_a = \sum_{a \in \arcsMod_i^-} \arcLoadMod_a, \hspace{0.5mm} \forall \hspace{0.5mm} i \ne o, d, \\ \nonumber
    &\hspace{1cm} \sum_{a \in \arcsMod_o^+} \arcLoadMod_a = \nodeLoadIn_o, \hspace{1mm} \arcLoadMod_a \geq 0, \hspace{0.5mm} \forall \hspace{0.5mm} a \in \arcsMod \Bigg\}
\end{align}

\subsection{Socially Optimal Tolls}
\label{subsec: Socially Optimal Tolls}

The objective of toll implementation is to realign commuter's incentives and route selection decisions, to induce \textit{perturbed social optimality} with respect to the logit Markovian model detailed in Section \ref{subsec: Setup}, as defined below.

\begin{definition}[\textbf{Perturbed Socially Optimal Flow}] \label{Def: Perturbed Socially Optimal Flow}
Let the perturbed total weighted latency $L: \arcsLoadConstraintSet \times \R^{|\arcsMod|} \times \R \ra \R$ be given by:
{\small
\begin{align} \label{Eqn: L, Overall Latency}
    &L(\arcLoadMod, \theta, \beta) \\ \nonumber
    := \hspace{0.5mm} &\sum_{a \in A} \arcLoadMod_a s_a(w_a)  + \\ \nonumber
    &\hspace{1mm} \frac{1}{\beta} \sum_{i \in \nodesMod \backslash \{d\}} \Bigg[ \sum_{a \in \arcsMod_i^+} \arcLoadMod_a \ln \arcLoadMod_a - \Bigg( \sum_{a \in \arcsMod_i^+} \arcLoadMod_a \Bigg) \ln\Bigg( \sum_{a \in \arcsMod_i^+} \arcLoadMod_a \Bigg) \Bigg].
\end{align}
}
We call $\arcLoadMod^\star \in \arcsLoadConstraintSet$ the \emph{perturbed socially optimal flow} with latency parameters $\theta$ and entropy parameter $\beta > 0$ if it solves $\min_{w \in \arcsLoadConstraintSet} L(\arcLoadMod, \theta, \beta)$, with $\arcsLoadConstraintSet$ given by \eqref{Eqn: Def, W}.
\end{definition}

In the perturbed total latency $L$ defined above, the first component is the total latency on the network weighted by the traffic load on each arc, while the second component is a non-positive entropy term that achieves its minimum when the traffic load at each non-destination node allocates itself equally among all outgoing arcs. Thus, the entropy parameter $\beta$ weights the total network latency against the tendency of commuters with imperfect information to explore among outgoing arcs at each intermediate node.

Since the minimization problem posed by Definition \ref{Def: Perturbed Socially Optimal Flow} is strictly convex, the perturbed socially optimal flow exists and is unique. Moreover, \cite{Chiu2023ArcbasedTrafficAssignment, Chiu2023DynamicTollinginArcBasedTAMs} establish that, given a traffic network $\graphMod = (\nodesMod, \arcsMod)$ with latency function parameters $\theta \in \R^{|\arcsMod|}$ and entropy parameter $\beta > 0$, there exists an \textit{optimal toll} $\bar \toll \in \R^{|\arcsMod|}$ whose corresponding MTE $\bar \arcLoadMod^{\theta, \beta}(\bar \toll)$ is perturbed socially optimal, and a dynamic tolling scheme that converges to the optimal toll. Those results, in the context of the online tolling problem considered in this work, are as summarized below. For more details, please see \cite{Chiu2023DynamicTollinginArcBasedTAMs}.

\begin{proposition} \label{Prop: Chiu et al}
There
exists $\tilde \arcLoadMod \in \arcsLoadConstraintSet$ and $\bar \toll \in \R^{|\arcsMod|}$ such that $\tilde \arcLoadMod = \bar \arcLoadMod^{\theta, \beta}(\bar \toll)$ and $\bar \toll_a^t = \bar \arcLoadMod_a^t \cdot \theta_a$ for each $a \in \arcsMod$. Moreover, $\bar \arcLoadMod$ is perturbed socially optimal, i.e., $\tilde \arcLoadMod = \emph{arg}\min_{\arcLoadMod \in \arcsLoadConstraintSet} L(\arcLoadMod, \theta, \beta)$.
\end{proposition}

\subsection{Online Learning Problem}
\label{subsec: Online Learning Problem}

Here, we pose the online learning problem that forms the central focus of this work. Let $T$ denote the total number of iterations for which the algorithm is run. Consider a traffic network $\graphMod$ with known node and arc structure $(\nodesMod, \arcsMod)$, but unknown latency function parameters $\{\theta_a^\star: a \in \arcsMod\}$ and entropy parameter $\beta^\star > 0$. We assume that $\theta^\star$ and $\beta^\star$ are bounded, as posed below. 

\begin{assumption}[\textbf{Parameter Bounds}]
There exist constants $c_\theta, C_\theta, c_\beta > 0$ such that $\theta_a^\star \in [c_\theta, C_\theta]$ for each $a \in \arcsMod$, and $\beta^\star > c_\beta$. The central authority has access to $c_\beta$ but not necessarily $c_\theta$ or $C_\theta$.
\end{assumption}

The above assumptions are not overly restrictive, since roads cannot be arbitrarily congestive, and travelers usually have some non-zero proclivity for selecting cost-minimizing arcs and routes. Moreover, as established in Section \ref{sec: Main Algorithm}, the arc latency parameter estimation errors $\Vert \theta^t - \theta^\star \Vert_2$ shrinks rapidly as $t$ increases. This allows the true, unknown temperature parameter $\beta^\star$, and thus a lower bound for $\beta^\star$, to be estimated with increasing accuracy as more data is collected.

Now, consider ourselves in the position of a central traffic authority that wishes to minimize the perturbed total latency over the iterations $t \in [T]$, despite initially lacking knowledge of the function parameters $\theta \in \R^{|\arcsMod|}$, and the underlying entropy parameter $\beta$. To accomplish this, at each iteration $t \in [T]$, we implement a toll vector $\hat p^t \in \R^{|\arcsMod|}$, and observe the resulting MTE traffic load allocation $\arcLoadMod^t := \bar \arcLoadMod^{\theta^\star, \beta^\star}(\toll^t) \in \arcsLoadConstraintSet$, as well as the random realizations of the travelers' latencies on each arc:
\begin{align*}
    \ell_{a,j}^t &= \latency_a(\arcLoadMod_a^t) + \epsilon_{a,j}^t.
\end{align*}
for each $j \in [\lfloor \arcLoadMod^t \rfloor]$, where $\epsilon_{a,j}^t$ are independent 1-subGaussian random variables. We then use the flow data $\{\arcLoadMod_a^t: a \in \arcsMod\}$ and the latency data $\{\ell_{a,j}^t: a \in \arcsMod, j \in [\lfloor \arcLoadMod^t \rfloor] \}$ to update our estimates of the underlying, unknown latency function  parameter $\theta^\star$ and entropy parameter $\beta^\star$, and correspondingly design our toll to implement at the next iteration $t+1$. The cumulative regret $R$ over the iterations $t \in [T]$ is thus given by:
{\small
\begin{align} \label{Def: Regret}
    R := \sum_{t=1}^T \big[ L\big( \bar \arcLoadMod^{\theta^\star, \beta^\star}(\toll^t), \theta^\star, \beta^\star \big) - L\big( \bar \arcLoadMod^{\theta^\star, \beta^\star}(\toll^\star), \theta^\star, \beta^\star \big) \big]
\end{align}
}

The core tenet of the above framework is that, as we accumulate more data on the traffic flow and realized latencies, we can construct increasingly accurate estimates of $\theta^\star$ and $\beta^\star$, and consequently adapt our tolls $\toll^t$ to reduce congestion in an increasingly effective manner.

\section{Main Algorithm}
\label{sec: Main Algorithm}

In this section, we present the main components of our algorithm (Algorithm \ref{Alg: Simultaneous Tolling and Parameter Estimation}). Section \ref{subsec: Least-Squares Estimator for Latency Function Parameters} describes the least-squares estimator used to approximate the arc latency functions from collected flow data. Section \ref{subsec: Entropy Parameter Estimation} then discusses our novel approximation scheme for the unknown entropy parameter $\beta$. Finally, we present our main algorithm in Section \ref{subsec: Algorithm Overview}.

\subsection{Least-Squares Estimator for Latency Function Parameters}
\label{subsec: Least-Squares Estimator for Latency Function Parameters}

First, we present the regularized least-squares estimator for the arc latency coefficients $\{\theta_a: a \in \arcsMod\}$. At each iteration $t \in [T]$, for each arc $a \in \arcsMod$, we observe the traffic flow at the current iteration,
$\arcLoadMod_a^t$, and latency data $\{\ell_{a,j}^k: a \in \arcsMod, k \in [t], j \in \lfloor \arcLoadMod_a^t \rfloor \}$, We then update the regularized least-squares estimate $\hat \theta_a^t > 0$ for the true coefficient $\theta_a^\star$, with regularizer $\lambda_a > 0$, as follows \footnote{We assume $\lfloor \arcLoadMod \rfloor \geq 1$, i.e., each arc is traversed upon by at least one commuter per iteration.}:
\begin{align*}
    \hat \theta_a^t := \text{arg} \min_{\theta_a} \left( \sum_{j=1}^{t-1} \sum_{k=1}^{\lfloor \arcLoadMod_a^j \rfloor} (\ell_{a,k}^j - \theta_a \arcLoadMod_a^j)^2 + \lambda_a \Vert \theta_a \Vert_2^2 \right).
\end{align*}
The following lemma states that these estimates, across iterations $t \in [T]$, lie within a neighborhood of the true parameter $\theta_a^\star$.

\begin{lemma} \label{Lemma: Good event probability, lower bound} \cite{LattimoreSzepesvari2020BanditAlgorithms}
For each $t \in [T]$ arc $a \in \arcsMod$, define:
\begin{align} \nonumber
    V_a^t &:= \sum_{\tau = 1}^t \lfloor w_a^t \rfloor (w_a^t)^2, \\ \label{Eqn: gamma}
    \gamma_a^t &:= \sqrt{\lambda_a} C_\theta + \sqrt{2 \ln T + 2 \ln\left( \frac{V_a^{t-1}}{\lambda_a} \right)},
\end{align}
and let the \say{good event} $E$ be defined by:
\begin{align*}
    E := \left\{ \forall \hspace{0.5mm} t \in [T], \forall \hspace{0.5mm} a \in \arcsMod: \hspace{0.5mm} |\hat \theta_a^{t-1} - \theta_a^\star| \leq \frac{\gamma_a^t}{\sqrt{V_a^{t-1}}} \right\}.
\end{align*}
Then $\Prob(E) \geq 1 - \frac{|\arcsMod|}{T}$.
\end{lemma}

\begin{proof}(\textbf{Sketch})
We construct upper confidence bounds for the least square estimator using covering arguments and martingale theory, as is standard in the bandit literature (see \cite{LattimoreSzepesvari2020BanditAlgorithms}, Chapter 20, and \cite{Gollapudi2023OnlineLearningforTrafficNavigation}.) For details, please see Appendix \ref{subsec: A1, Proof of Lemma: Good event probability, lower bound}.
\end{proof}

In words, with probability at least $1 - \frac{|\arcsMod|}{T}$, for each arc $a \in \arcsMod$ at each iteration $t \in [T]$, the estimate $\hat \theta_a^t$ falls within the confidence interval $\Big[ \theta_a^\star - \frac{\gamma_a^t}{\sqrt{V_a^{t-1}}}, \theta_a^\star +  \frac{\gamma_a^t}{\sqrt{V_a^{t-1}}} \Big]$. Below, for convenience, we set:
\begin{align*}
    \hat \theta_a^{t,-} &:= \hat \theta_a^t - \frac{\gamma_a^t}{\sqrt{V_a^{t-1}}}, \\
    \hat \theta_a^{t,+} &:= \hat \theta_a^t +  \frac{\gamma_a^t}{\sqrt{V_a^{t-1}}} 
\end{align*}

\subsection{Entropy Parameter Estimation}
\label{subsec: Entropy Parameter Estimation}

Intuitively, the entropy parameter governs the degree to which travelers at an intermediate node prefer to select an outgoing arc that minimizes the cost-to-go. Specifically, when $\beta \ra \infty$, travelers at node $i$ select with probability 1 an outgoing arc $a \in \arcsMod_i^+$ that minimizes the cost-to-go; when $\beta \ra 0$, travelers at node $i$ select from all outgoing arcs with equal probability, essentially ignoring their cost-to-go values. As such, a natural approach for estimate $\beta$ would begin by fixing a node $i^\star$, whose outgoing routes to the destination are relatively straightforward to describe. Then, we can analyze data that characterize the traffic flows and costs among its outgoing arcs at each iteration $t \in [T]$, to gain insight into the strength of the commuters' preference to minimize their cost-to-go, i.e., to estimate $\beta$.

\begin{figure}
    \centering
    \includegraphics[scale=0.12]{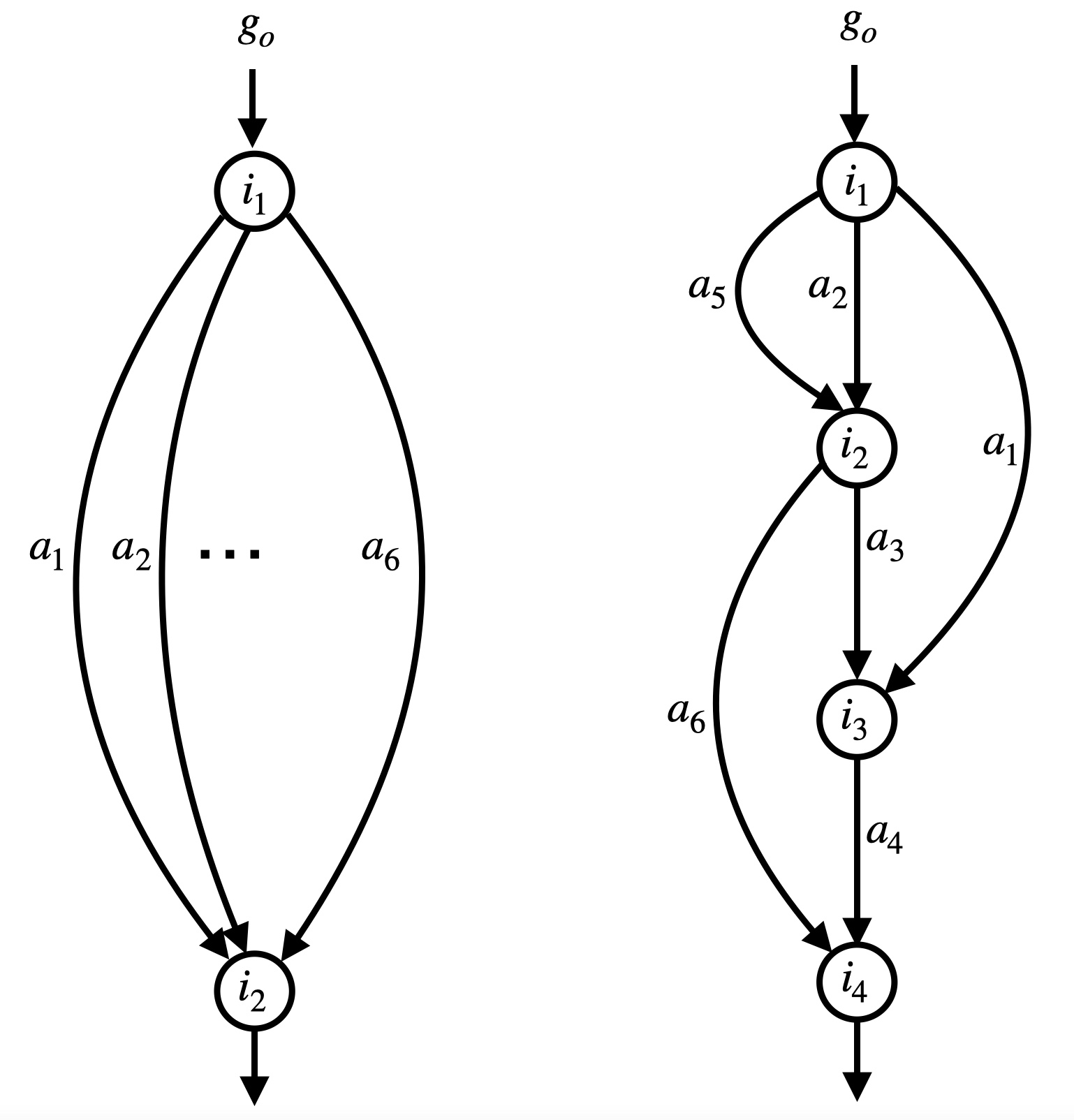}
    \caption{(Left) A parallel 6-arc network; here, $i^\star = i_1$. (Right) A more general network with 6 arcs; here, $i^\star = i_2$, since there are two routes from $i_2$ to the destination $i_4$ which do not share an arc.}
    \label{fig:Network Schematics}
\end{figure}

We thus begin with the following lemma, which states that regardless of the precise structure of the traffic network $\graphMod$, there must exist a node $i^\star \in \nodesMod$ with properties desirable for estimating $\beta > 0$. For every node $i^\star$ satisfying the conditions of Lemma \ref{Lemma: Existence of Node for Estimating beta}, each outgoing arc $a' \in \arcsMod_{i^\star}^+$ yields exactly one route from $i^\star$ to $d$. Thus, the route segments from $i^\star$ to $d$ have structure akin to a parallel-link network, allowing the estimation of the entropy parameter $\beta^\star$ from $i^\star$ to be straightforward. Examples are furnished in Figure \ref{fig:Network Schematics}.

\begin{lemma} \label{Lemma: Existence of Node for Estimating beta}
There exists a node $i^\star \in \nodesMod \backslash \{d\}$ such that $|\arcsMod_{i^\star}^+| \geq 2$, and for each $j \in \arcsMod_{i^\star}^+$, either $j = d$, or there exists only one route from $j$ to $d$.
\end{lemma}

\begin{proof}(\textbf{Sketch})
This follows by starting from the destination $d$ and recursively searching for the desired node $i^\star$ by moving back towards the origin $o$. For details, please see Appendix \ref{subsec: A1, Proof of Lemma: Existence of Node for Estimating beta}.
\end{proof}

Below, we present assumptions that facilitate the estimation of the true, unknown temperature parameter $\beta^\star$. First, for each node $i^\star \in \nodesMod \backslash \{d\}$, and any arc latency parameter estimate $\theta \in \R^{|\arcsMod|}$ and temperature parameter $\beta > 0$ within a range of reasonable estimates for the true parameters $\theta^\star \in \R^{|\arcsMod|}$ and $\beta^\star > 0$, we assume that the MTE costs of the outgoing edges $\arcsMod_{i^\star}^+$ are not identical. In particular, for each such node $i^\star$, among the outgoing arcs $\arcsMod$, there must be sufficiently differentiation, in the form of a strictly positive gap $\Delta_z > 0$, between the minimum and maximum costs-to-go. This facilitates the estimation of the temperature parameter in $\beta$, and emphasizes its role in the stochastic route choices made on the part of the travelers. Indeed, the temperature parameter $\beta$ is not meaningful in networks with route segments that are virtually indistinguishable in cost.

\begin{assumption} \label{Assumption: Delta z}
Let $\bar \toll(\hat \theta, \hat \beta)$ denote the optimal toll corresponding to an arc-based TAM with entropy parameter $\hat \beta$, over a network with latency function parameters $\theta$. There exists $\Delta_z > 0$, such that, for any node $i^\star \in \nodesMod$ satisfying the conditions of Lemma \ref{Lemma: Existence of Node for Estimating beta}, and any parameter estimates within known bounds, $\hat \theta \in [c_\theta, C_\theta]$ and $\hat \beta \in [c_\beta, \infty)$, we have:
    \begin{align*}
        &\max_{a' \in \arcsMod_{i^\star}^+} z_{a'}(\hat \theta, \bar \arcLoadMod^{\theta^\star, \beta^\star}(\bar \toll(\hat \theta, \hat \beta)), \bar \toll(\hat \theta, \hat \beta)) \\
        &\hspace{1cm} - \min_{a' \in \arcsMod_{i^\star}^+} z_{a'}(\theta, \bar \arcLoadMod^{\theta^\star, \beta^\star}(\bar \toll(\hat \theta, \hat \beta)), \bar \toll(\hat \theta, \hat \beta)) \geq \Delta_z.
    \end{align*}
\end{assumption}

In the following lemma, we establish an estimator $\beta^t$ for the temperature parameter $\beta$ at each iteration $t$ whose proximity to the true temperature parameter $\beta^\star$ is directly proportional to the gap between the under- and over-estimators $\theta^{t,-} \in \R^{|\arcsMod|}$ and $\theta^{t,+} \in \R^{|\arcsMod|}$ of the true arc latency parameter $\theta^\star$. The key intuition behind the estimator is that, if the true latency function parameters $\theta^\star$ on each arc were known, the underlying entropy parameter $\beta^\star$ can be perfectly recovered by comparing the flows of outgoing arcs at a non-destination node, and the ratios between the costs-to-go of these arcs. However, since the central authority lacks access to $\theta^\star$, we instead use the upper and lower bounds of the confidence interval at each iteration $t$, i.e., $\{\theta_a^{t,+}, \theta_a^{t,-}: a \in \arcsMod\}$, to construct an estimate $\beta^t$ of the underlying, unknown entropy parameter $\beta^\star$. Moreover, we construct the estimate $\beta^t$ to provably under-approximate $\beta^\star$, i.e.,  to guarantee that $\beta^t \leq \beta^\star$. This can be viewed as an extension of the Principle of Optimism in the Face of Uncertainty, since the total latency \eqref{Eqn: L, Overall Latency} is non-decreasing in the entropy parameter $\beta$ (Recall that the entropy term, to which the $1/\beta^\star$ factor is multiplied, is always non-positive).

\begin{lemma} \label{Lemma: Beta Estimator}
Let $i^\star \in \nodesMod \backslash \{d\}$ be any node satisfying the conditions in Lemma \ref{Lemma: Existence of Node for Estimating beta}, and let:
\begin{align}
    a^\star \in \text{arg} \min_{a' \in \arcsMod_{i^\star}^+} z_{a'}(\theta^\star, \arcLoadMod^t, \toll^t).
\end{align}
Then there exists $\beta^t \in [c_\beta, \beta^\star]$ such that:
\begin{align} \label{Eqn: Fixed Point Equation}
    &\frac{\exp(-\beta^t \cdot z_{a^\star}(\theta^{t,-}, \arcLoadMod^t, \toll^t))}{\sum_{a' \in \arcsMod_{i^\star}^+} \exp(-\beta^t \cdot z_{a'}(\theta^{t,+}, \arcLoadMod^t, \toll^t))} \\ \nonumber
    = \hspace{0.5mm} &\frac{\exp(-\beta^\star \cdot z_{a^\star}(\theta^\star, \arcLoadMod^t, \toll^t))}{\sum_{a' \in \arcsMod_{i^\star}^+} \exp(-\beta^\star \cdot z_{a'}(\theta^\star, \arcLoadMod^t, \toll^t))}
\end{align}
Moreover, let $\arcsMod(i^\star)$ denote the set of all arcs contained in a route from $i^\star$ to $d$. Then:
\begin{align} \label{Eqn: Upper Bound for beta estimate, using theta estimate}
    |\beta^t - \beta^\star| \leq \frac{\beta^\star g_o}{\Delta_z} \cdot \sum_{a \in \arcsMod(i^\star)} (\theta_a^{t,+} - \theta_a^{t,-}) \arcLoadMod_a^t.
\end{align}
\end{lemma}

\begin{proof}(\textbf{Sketch})
Lemma \ref{Lemma: Beta Estimator} follows from the monotonicity of the exponential function, as well as the monotonicity of the cost-to-go terms $z_a$ with respect to the latency function parameters $\theta$. For details, please see Appendix \ref{subsec: A1, Proof of Lemma: Beta Estimator}.
\end{proof}

The upper bound \eqref{Eqn: Upper Bound for beta estimate, using theta estimate} demonstrates that, by applying the least-squares estimator described in Section \ref{subsec: Least-Squares Estimator for Latency Function Parameters}, which ensures that $\Vert \theta^{t,+} - \theta^{t,-} \Vert_2 < O(1/\sqrt{t})$ as $t \ra \infty$, we can likewise ensure that $|\beta^t - \beta^\star| < O(1/\sqrt{t})$ as $t \ra \infty$.

\subsection{Algorithm Overview}
\label{subsec: Algorithm Overview}

Armed with the estimation schemes for $\theta^\star$ and $\beta^\star$ presented in Sections \ref{subsec: Least-Squares Estimator for Latency Function Parameters} and \ref{subsec: Entropy Parameter Estimation}, we proceed to present our online learning algorithm (Algorithm \ref{Alg: Simultaneous Tolling and Parameter Estimation}). At each iteration $t$, the central authority uses latency function and entropy parameter estimates obtained in the previous round to compute the corresponding optimal toll $\toll^t$ (Line \ref{Eqn: Alg 1, p t}). Observe that, for the latency function parameter, we use the lower bound $\theta^{t,-}$ of the confidence interval $(\theta^{t,-}, \theta^{t,+})$, in accordance with the Principle of Optimism in the Face of Uncertainty. Commuters then sequentially select arcs in the traffic network to minimize their average cost-to-go, resulting in the MTE traffic allocation $\arcLoadMod^t := \bar \arcLoadMod^{\theta^\star, \beta^\star}(\toll^t)$ (Line \ref{Eqn: Alg 1, w t}). The central authority then collects this data, and uses the regularized least-squares method in Section \ref{subsec: Least-Squares Estimator for Latency Function Parameters} to construct an updated estimate $\theta^t$ of the underlying latency function parameters $\theta^\star$ (Lines \ref{Eqn: Alg 1, theta estimation, Start}-\ref{Eqn: Alg 1, theta estimation, End}). Finally, we construct an update estimate $\beta^t$ of the underlying entropy parameter $\beta^\star$ using the approach in Section \ref{subsec: Entropy Parameter Estimation} (Lines \ref{Eqn: Alg 1, beta estimation, Start}-\ref{Eqn: Alg 1, beta estimation, End}).

\begin{algorithm} 
{
\normalsize
\SetAlgoLined

\KwData{$i^\star \in \nodesMod$, $\beta^0 := c_\beta > 0$, $\lambda_a$, $V_a^0 = \lambda_a$, $Q_a^0 = 0$, and $\toll_a^0$ $\theta_a^{0,-} > 0$, $\theta_a^{0,+} > 0$ ($\forall a \in \arcsMod$)}

 \vspace{2mm}
 
 \For{$t = 1, \cdots, T$}{

    \For{$a \in \arcsMod$}{
        $\ell_{a,1}^t, \cdots, \ell_{a,\lfloor \arcLoadMod_a^{t-1} \rfloor}^t \gets$ Costs collected from arc $a$ at iteration $t$ \label{Eqn: Alg 1, theta estimation, Start}

        $\gamma_a^t \gets \sqrt{\lambda_a} C_\theta + \sqrt{2 \ln T + \ln\Big( \frac{V_a^{t-1}}{\lambda_a} \Big)}$

        $V_a^t \gets V_a^{t-1} + \lfloor \arcLoadMod_a^{t-1} \rfloor (\arcLoadMod_a^{t-1})^2$ 

        $Q_a^t \gets Q_a^{t-1} + \arcLoadMod_a^{t-1} \cdot \sum_{k=1}^{\lfloor \arcLoadMod_a^{t-1} \rfloor} \ell_{a, k}^t$

        $\hat \theta_a^t \gets Q_a^t / V_a^t$
        \label{Eqn: Alg 1, theta estimation, End}

        $\theta_a^{t,-} \gets \max\Big\{ \hat \theta_a^{t-1} - \frac{\gamma_a^{t}}{\sqrt{V_a^{t-1}}}, 0 \Big\}$

        $\theta_a^{t,+} \gets \hat \theta_a^{t-1} + \frac{\gamma_a^{t}}{\sqrt{V_a^{t-1}}}$ 
    }
    
    $\tilde \beta^{t} \gets$ Solution to---$\forall a \in \arcsMod_{i^\star}^+$:
    \begin{align*}
        \frac{\arcLoadMod_a^t}{\sum\limits_{a' \in  \arcsMod_{i^\star}^+} \arcLoadMod_{a'}^t} &= \frac{\exp(-\tilde \beta^{t} \cdot \costToGo_a(\theta^{t, -}, \arcLoadMod^t, \toll^t))}{\sum\limits_{a' \in \arcsMod_{i^\star}^+} \exp(-\tilde \beta^{t} \cdot \costToGo_a(\theta^{t, +}, \arcLoadMod^t, \toll^t))}.
    \end{align*}
    \label{Eqn: Alg 1, beta estimation, Start}

    $\beta^{t} \gets \max\{c_\beta, \tilde \beta^{t}\}$. \label{Eqn: Alg 1, beta estimation, End}

    $\toll^t \gets$ Solution to 
    $\toll = \theta^{t-1,-} \cdot \bar \arcLoadMod^{\theta^{t-1,-}, \beta^t}(\toll)$. \label{Eqn: Alg 1, p t}

    $\arcLoadMod^t \gets \bar \arcLoadMod^{\theta^\star, \beta^\star}(\toll^t)$ (Commuters' flow allocation) \label{Eqn: Alg 1, w t}
 }
 \caption{Simultaneous Tolling and Parameter Estimation}
 \label{Alg: Simultaneous Tolling and Parameter Estimation}
 }
\end{algorithm}

\section{Regret Analysis}
\label{sec: Regret Analysis}

Here, we upper bound the regret incurred by Algorithm \ref{Alg: Simultaneous Tolling and Parameter Estimation}. First, we require the following lemma, which facilitates the decomposition of the regret into tractable terms.

\begin{lemma} \label{Lemma: Cost, Monotonicity}
Suppose $\theta_a^1 \leq \theta_a^2$ for each $a \in \arcsMod$, and $\beta^1 \leq \beta^2$. Then, for each $\arcLoadMod \in \arcsLoadConstraintSet$:
\begin{align*}
    L(\arcLoadMod, \theta^1, \beta^1) \leq L(\arcLoadMod, \theta^2, \beta^2).
\end{align*}
\end{lemma}

\begin{proof}
This follows by noting that $\arcLoadMod \geq 0$, and that the entropy term in $L$ is non-positive.
\end{proof}

We now present our regret bound.

\begin{theorem} \label{Thm: Regret Analysis}
There exists $K(\lambda, \Delta_z, c_\theta, C_\theta, c_\beta, \beta^\star) > 0$ such that for any $T \in \N$:
\begin{align*}
    R \leq K g_o^2 \ln^2(g_o) |\arcsMod| \sqrt{T} \ln(T g_o) \max\left\{|\nodesMod| \ln\left( \frac{|\arcsMod|}{|\nodesMod|} \right), B \right\},
\end{align*}
where $B := |\arcsMod(i^\star)|$ denotes the set of all arcs used to construct the estimates $\beta^t$.
\end{theorem}

\begin{proof}(\textbf{Proof Sketch})
As in Algorithm \ref{Alg: Simultaneous Tolling and Parameter Estimation}, set $\toll^t \in \R^{|\arcsMod|}$ and $\toll^\star \in \R^{|\arcsMod|}$ to be the unique solutions to the following fixed-point equations:
\begin{align*}
    \toll^t &= \theta^{t-1,-} \cdot \bar \arcLoadMod^{\theta^{t-1}, \beta^{t-1}}(\toll^t), \\
    \toll^\star &= \theta^\star \cdot \bar \arcLoadMod^{\theta^\star, \beta^\star}(\toll^\star).
\end{align*}

Under the good event $E$ described in Lemma \ref{Lemma: Good event probability, lower bound}:
\begin{align*}
    \latencyTotal\big( \bar \arcLoadMod^{\theta^{t,-}, \beta^t}(\toll^t), \theta^{t,-}, \beta^t \big) &\leq \latencyTotal\big( \bar \arcLoadMod^{\theta^\star, \beta^\star}(\toll^\star), \theta^{t,-}, \beta^t \big) \\
    &\leq \latencyTotal\big( \bar \arcLoadMod^{\theta^\star, \beta^\star}(\toll^\star), \theta^\star, \beta^\star \big),
\end{align*}
where the first inequality follows since Definition \ref{Def: Perturbed Socially Optimal Flow}, Proposition \ref{Prop: Chiu et al}, and the definition of $\toll^t$ (Algorithm \ref{Alg: Simultaneous Tolling and Parameter Estimation}, Line \ref{Eqn: Alg 1, p t}) together imply that $\bar \arcLoadMod^{\theta^{t,-}, \beta^t}(\toll^t) = \text{arg}\min_{\arcLoadMod \in \arcsLoadConstraintSet} L(\arcLoadMod, \theta^{t,-}, \beta^\star, \beta^t)$, while the second inequality follows from Lemmas \ref{Lemma: Good event probability, lower bound} and \ref{Lemma: Cost, Monotonicity}. 

Define $\chi: \arcsLoadConstraintSet \ra \R$ to be the entropy term in $C$:
{\small
\begin{align} \label{Eqn: chi, entropy term}
    &\chi(\arcLoadMod) \\ \nonumber
    := \hspace{0.5mm} &\sum_{i \in \nodesMod \backslash \{d\}} \left[ \sum_{a \in \arcsMod_i^+} \arcLoadMod_a \ln \arcLoadMod_a - \left( \sum_{a \in \arcsMod_i^+} \arcLoadMod_a \right) \ln \left( \sum_{a \in \arcsMod_i^+} \arcLoadMod_a \right) \right]
\end{align}
}
Thus, the regret $R$ can be upper bounded as follows:
{\small
\begin{align} \nonumber
    R &= \sum_{t=1}^T \big[ \latencyTotal\big( \bar \arcLoadMod^{\theta^\star, \beta^\star}(\toll^t), \theta^\star, \beta^\star \big) - \latencyTotal\big( \bar \arcLoadMod^{\theta^\star, \beta^\star}(\toll^\star), \theta^\star, \beta^\star \big) \big] \\ \nonumber
    &\leq \sum_{t=1}^T \big[ \latencyTotal\big( \bar \arcLoadMod^{\theta^\star, \beta^\star}(\toll^t), \theta^\star, \beta^\star \big) - \latencyTotal\big( \bar \arcLoadMod^{\theta^{t,-}, \beta^t}(\toll^t), \theta^{t,-}, \beta^t \big) \big] \\ \nonumber
    &= \sum_{t=1}^T \big[ \latencyTotal\big( \bar \arcLoadMod^{\theta^\star, \beta^\star}(\toll^t), \theta^\star, \beta^\star \big) - \latencyTotal\big( \bar \arcLoadMod^{\theta^\star, \beta^\star}(\toll^t), \theta^{t,-}, \beta^t \big) \big] \\ \nonumber
    &\hspace{5mm} + \sum_{t=1}^T \big[ \latencyTotal\big( \bar \arcLoadMod^{\theta^\star, \beta^\star}(\toll^t), \theta^{t,-}, \beta^t \big) \\ \nonumber
    &\hspace{1.5cm} - \latencyTotal\big( \bar \arcLoadMod^{\theta^{t,-}, \beta^t}(\toll^t), \theta^{t,-}, \beta^t \big) \big] \\ \label{Eqn: R1}
    &= \sum_{t=1}^T \sum_{a \in \arcsMod} (\theta_a^\star - \theta_a^{t,-}) \bar \arcLoadMod_a^{\theta^\star, \beta^\star}(\toll^t)^2 \\ \label{Eqn: R2}
    &\hspace{5mm} + \sum_{t=1}^T \left( \frac{1}{\beta^\star} - \frac{1}{\beta^t} \right) \cdot \chi\big( \bar \arcLoadMod_a^{\theta^\star, \beta^\star}(\toll^t) \big) \\ \label{Eqn: R3}
    &\hspace{5mm} + \sum_{t=1}^T \big[ \latencyTotal\big( \bar \arcLoadMod^{\theta^\star, \beta^\star}(\toll^t), \theta^{t,-}, \beta^t \big) \\ \nonumber
    &\hspace{1.5cm} - \latencyTotal\big( \bar \arcLoadMod^{\theta^{t,-}, \beta^t}(\toll^t), \theta^{t,-}, \beta^t \big) \big],
\end{align}
}
where, in accordance with the notation in Algorithm \ref{Alg: Simultaneous Tolling and Parameter Estimation}, we set $\arcLoadMod^t := \bar \arcLoadMod^{\theta^\star, \beta^\star}(\toll^t)$. Define the three summands \eqref{Eqn: R1}, \eqref{Eqn: R2}, \eqref{Eqn: R3} by $R_1$, $R_2$, and $R_3$ respectively. The convergence rate of $\theta^{t,-} \ra \theta^\star$ and $\beta^t \ra \beta^\star$ can then be analyzed to yield non-asymptotic bounds for $R_1$ and $R_2$, respectively. In turn, these bounds are then used to bound $R_3$.

For more details, please see Appendices \ref{subsec: A2, Upper Bound for R1}, \ref{subsec: A2, Upper Bound for R2}, and \ref{subsec: A2, Upper Bound for R3}.
\end{proof}

\begin{remark} \label{Remark: Regret Upper Bound}
Compared to \cite{Gollapudi2023OnlineLearningforTrafficNavigation}, our regret upper bound contains an extra term $\max\{|\nodesMod| \ln(|\arcsMod|/|\nodesMod|), B\}$, due to the following unique features of our problem formulation: (1) Entropy parameter estimation, which contributes the network structure-dependent constant $B$, (2) The tolling authority affects the equilibrium flow allocation indirectly, through tolls, instead of directly dictating commuters' route selections, (3) Mismatch between the latency function and entropy parameter estimates $(\theta^{t,-}, \beta^t)$ used by the tolling authority to compute tolls, and the true parameters $(\theta^\star, \beta^\star)$ used by the commuters to best-respond to the implemented toll.
\end{remark}

\section{Experiments}
\label{sec: Experiments}

We present numerical results on simulated traffic networks that validate the regret bounds presented in Theorem \ref{Thm: Regret Analysis}. We ran Algorithm \ref{Alg: Simultaneous Tolling and Parameter Estimation} for $T = 2500$ iterations, with $g_o = 100$, on the parallel-arc network in Figure \ref{fig:Network Schematics} (left), with underlying parameters $\theta^\star := (1.5, 2.5, 3.5, 4.5, 5.5, 6.5) \in \R^6$, and $\beta^\star = 0.25$, and on the more general network in Figure \ref{fig:Network Schematics} (right), with underlying parameters $\theta^\star := (0.6, 0.4, 0.4, 0.4, 0.6, 0.6) \in \R^6$, and $\beta^\star = 0.25$. To suppress constants in the cumulative regret, we selected $\lambda_a = 0.01$ for each $a \in [6]$. For convenience, for each iteration $t \in [T]$, let $L^t := L(\arcLoadMod^{\theta^\star, \beta^\star}(\toll^t), \theta^\star, \beta^\star)$ denote the cost incurred at iteration $t$, let $L^\star := L\big( \arcLoadMod^{\theta^\star, \beta^\star}(\toll^\star), \theta^\star, \beta^\star \big)$ denote the minimum possible cost, and let $R^t := \sum_{\tau=1}^t \big[ L\big( \arcLoadMod^{\theta^\star, \beta^\star}(\toll^\tau), \theta^\star, \beta^\star \big) - L\big( \arcLoadMod^{\theta^\star, \beta^\star}(\toll^\star), \theta^\star, \beta^\star \big) \big]$ denote the cumulative regret up to iteration $t$. 

\begin{figure*}[ht]
    \centering
    \includegraphics[scale=0.19]{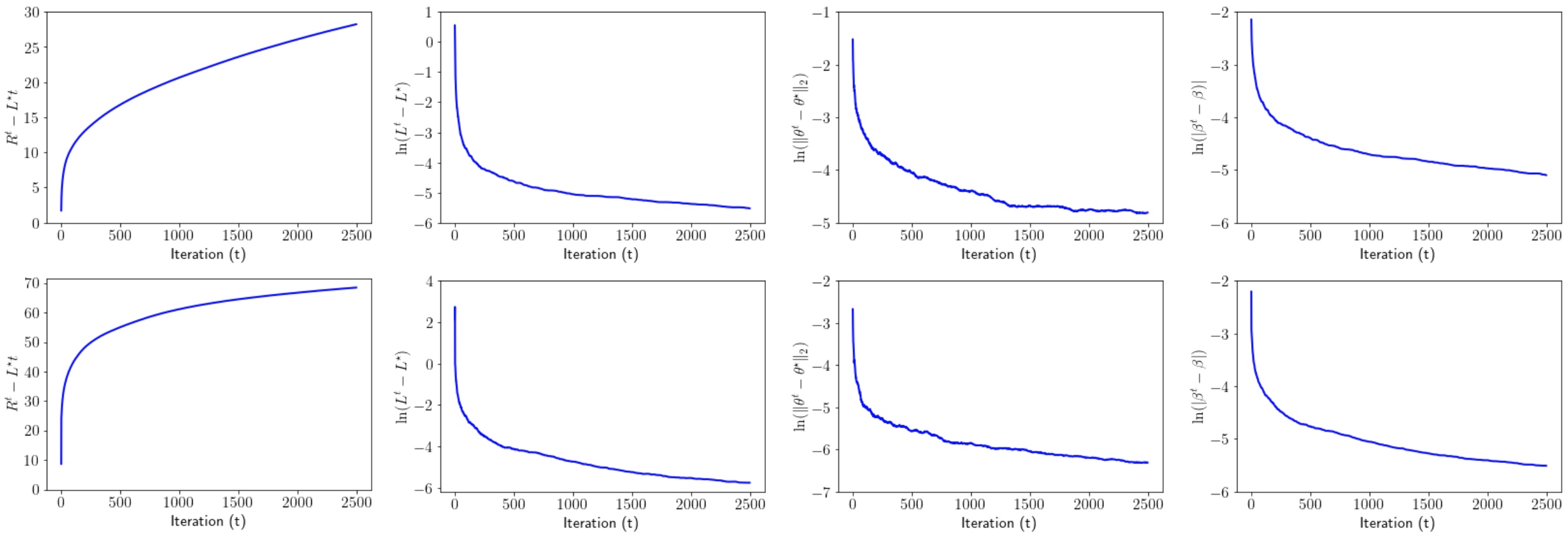}
    \caption{(Left to right) The cumulative regret $R^t - L^\star t$, logarithm of stage-wise regret $\ln(L^t - L^\star)$, logarithm of $\theta$-estimation error $\ln(\Vert \theta^t - \theta^\star \Vert_2)$, and logarithm of stage-wise regret $\ln(|\beta^t - \beta^\star|)$ for the parallel-arc network in Figure \ref{fig:Network Schematics} (top) and the more general network in Figure \ref{fig:Network Schematics} (bottom), as a function of the iteration count $t$. Note the sub-linear growth of the cumulative regret with respect to the iteration count, and the rapid decay of the stage-wise regret, $\theta$-estimation error, and $\beta$-estimation error to 0.}
    \label{fig:Results Figure}
\end{figure*}

Figure \ref{fig:Results Figure} illustrates the growth of the cumulative regret $R^t - L^\star t$ as a function of the iteration count $t$. We also provide logarithmic plots that describe the decay of the stage-wise regret $L^t - L^\star$, the magnitude of the latency function parameter estimation error $\Vert \theta \Vert_2$, and the magnitude of the entropy parameter estimation error $|\beta^t - \beta|$. For both networks, the cumulative regret increases as a sub-linear function of $t$, while the cumulative regret, $\theta$ estimation error, and $\beta$ estimation error decrease gracefully to 0 as $t$ increases.


\section{Conclusion and Future Work}
\label{sec: Conclusion and Future Work}

This work presents a novel online learning algorithm to learn the latency function and entropy parameters that characterize commuters' arc-selection decisions on a single source-single destination traffic network, while simultaneously implementing tolls to minimize the overall network congestion. We characterize a notion of regret using the accumulation across iterations of the gap between the incurred and minimum costs, and prove that our cumulative regret metric increases sub-linearly in the number of iterations $t$. Finally, we present numerical results illustrating the performance of our regret algorithm on simulated traffic networks.

A natural avenue of future work is to extend the results presented in this paper to traffic networks with multiple origin-destination pairs, and possibly bi-directional edges. Such settings pose particular challenges to the estimation of the entropy parameters, since each arc in the network could be shared among commuters with different travel histories and destinations. It would also be interesting to explore the relaxation of the assumption that the central authority possesses knowledge of a lower bound $c_\beta > 0$ for $\beta^\star$. 


\section{Acknowledgements} The authors would like to thank Chinmay Maheshwari and Pan-Yang Su for fruitful discussions regarding the arc-based congestion game formulation considered in this work.

\printbibliography



\newpage

\appendix

Below, we present proofs omitted in the main paper due to space limitations.

\section{Proofs for Section \ref{sec: Main Algorithm}}
\label{sec: A1, Main Algorithm}

First, we recall the definitions of the \textit{depth} and \textit{height} of a graph, as defined in \cite{Chiu2023ArcbasedTrafficAssignment} Appendix A, and restated below for completeness.

\begin{definition}[\textbf{Depth of a DAG}]
Given a DAG $\graphMod = (\nodesMod, \arcsMod)$ describing a single-origin single-destination traffic network, the \textit{depth of $\graphMod$}, denoted $\depth(G)$, is defined by: 
\begin{align*}
    \depth(\graphMod) := \max_{a \in \arcsMod} \depth_a
\end{align*}
\end{definition}

Since the acyclic traffic graphs studied in this work have finitely many edges, we have $\depth(G) < \infty$. Below, we summarize properties of the depth of a DAG.

\begin{proposition} \label{Prop: Condensed DAG Properties, Depth}
Given a Condensed DAG $\graphMod = (\nodesMod, \arcsMod)$ with the route set $\routes$:
\begin{enumerate}
    \item For any $a \in \arcsMod$, we have $\depth_a = 1$ if and only if $i_a = o$. Similarly, if $\depth_a = \depth(\graphMod)$, then $j_a = d$.
    
    \item For any fixed $r \in \routes$, and any $a, a' \in r$ with $\depth_{a, r} < \depth_{a',r}$, we have $\depth_a < \depth_{a'}$ i.e., arcs along a route have strictly increasing depth from the origin to the destination.
    
    \item Fix any $a \in \arcsMod$, and any $r \in \routes$ containing $a$ such that $\depth_{a,r} = \depth_a$. Then, for any $a' \in \routes$ preceding $a$ in $r$, we have $\depth_{a',r} = \depth_{a'}$.
    
    \item For each depth $k \in [\depth(G)] := \{1, \cdots, \depth(G)\}$, there exists some $a \in A$ such that $\depth_a = k$.
\end{enumerate}
\end{proposition}

\begin{proof}
See \cite{Chiu2023ArcbasedTrafficAssignment}, Appendix A.
\end{proof}

Similarly, we can define and characterize the height of a DAG.

\begin{definition}[\textbf{Height of a DAG}]
Given a DAG $\graphMod = (\nodesMod, \arcsMod)$ describing a single-origin single-destination traffic network, the \textit{height of $\graphMod$}, denoted $\height(G)$, is defined by: 
\begin{align*}
    \height(\graphMod) := \max_{a \in \arcsMod} \height_a
\end{align*}
\end{definition}

As with depth, we note that DAGs with finitely many edges have finite height, i.e., $\height(G) < \infty$.

\begin{proposition} \label{Prop: Condensed DAG Properties, Height}
Given an Condensed DAG $\graphMod = (\nodesMod, \arcsMod)$ with the route set $\routes$:
\begin{enumerate}
    \item For any $a \in \arcsMod$, we have $\height_a = 1$ if and only if $j_a = d$. Similarly, if $\height_a =\height(G)$, then $i_a = o$.
    
    \item For any fixed $r \in \routes$, and any $a, a' \in r$ with $\height_{a, r} < \height_{a',r}$, we have $\height_a < \height_{a'}$ i.e., arcs along a route from the origin to the destination have strictly decreasing depth.
    
    \item Fix any $a \in \arcsMod$, and any $r \in \routes$ containing $a$ such that $\height_{a,r} = \height_a$. Then, for any $a' \in \routes$ following $a$ in $r$, we have $\height_{a',r} = \height_{a'}$.
    
    \item For each height $k \in [\height(\graphMod)] := \{1, \cdots, \height(\graphMod)\}$, there exists an arc $a \in \arcsMod$ such that $\height_a = k$.
\end{enumerate}
\end{proposition}

\begin{proof}
See \cite{Chiu2023ArcbasedTrafficAssignment}, Appendix A.
\end{proof}

\subsection{Proof of Lemma \ref{Lemma: Good event probability, lower bound}}
\label{subsec: A1, Proof of Lemma: Good event probability, lower bound}

At each iteration $t \in [T]$, for each arc $a \in \arcsMod$, the regularized least-squares estimate $\hat \theta_a^t > 0$ for the true coefficient $\theta_a^\star$, with regularizer $\lambda_a > 0$, is given by:
\begin{align*}
    \hat \theta_a^t := \text{arg} \min_{\theta_a} \left( \sum_{j=1}^{t-1} \sum_{k=1}^{\lfloor \arcLoadMod_a^j \rfloor} (\depth_{a,k}^j - \theta_a \arcLoadMod_a^j)^2 + \lambda_a \Vert \theta_a \Vert_2^2 \right).
\end{align*}
Note that the cost objective in the above argmin expression is convex and quadratic. Thus, by setting the gradient to 0, we can compute the optimal parameter estimate as follows (for more details, please see Gollapudi et al. \cite{Gollapudi2023OnlineLearningforTrafficNavigation}, Lemma 2):
\begin{align} \label{Eqn: hat theta expression, 1}
    \hat \theta_a^t = \left( \lambda_a + \sum_{j=1}^{t-1} (\arcLoadMod_a^j)^3 \right)^{-1} \left( \sum_{j=1}^{t-1} \arcLoadMod_a^j \cdot \sum_{k=1}^{\lfloor \arcLoadMod_a^j \rfloor} \depth_{a,k}^j \right)
\end{align}
For convenience, we define:
\begin{align} \label{Eqn: V}
    V_a^t &:= \lambda_a + \sum_{j=1}^{t-1} (\arcLoadMod_a^j)^3, \\ \label{Eqn: W}
    W_a^t &:= \sum_{j=1}^{t-1} (\arcLoadMod_a^j)^3, \\ \label{Eqn: U}
    U_a^t &:= \sum_{j=1}^{t-1} \arcLoadMod_a^j \cdot \sum_{k=1}^{\lfloor \arcLoadMod_a^j \rfloor} \depth_{a,k}^j, \\ \label{Eqn: S}
    S_a^t &:= \sum_{j=1}^{t-1} \arcLoadMod_a^j \cdot \sum_{k=1}^{\lfloor \arcLoadMod_a^j \rfloor} \epsilon_{a,k}^j.
\end{align}
Thus, we can write \eqref{Eqn: hat theta expression, 1} as:
\begin{align} \label{Eqn: hat theta expression, 2}
    \hat \theta_a^t &= (V_a^t)^{-1} U_a^t = (V_a^t)^{-1} (W_a^t \theta_a + S_a^t).
\end{align}

For each arc $a \in \arcsMod$, the above process generates regularized least-squares estimates $\{\hat \theta_a^t\}$, across iterations $t \in [T]$, for the true underlying parameter $\theta_a^\star$. The following lemma demonstrates that these estimates, across iterations $t \in [T]$, lie within a neighborhood of the true parameter $\theta_a^\star$.


\begin{proof}(\textbf{Proof of Lemma \ref{Lemma: Good event probability, lower bound}})
The following proof parallels that of Gollapudi et al. \cite{Gollapudi2023OnlineLearningforTrafficNavigation}, Lemma 3, and is included for completeness.

From \eqref{Eqn: hat theta expression, 2}, we have:
\begin{align} \nonumber
    &\hspace{5mm} \sqrt{V_a^t} |\hat \theta_a^t - \theta_a^\star| \\ \nonumber
    &= \sqrt{V_a^t} |(V_a^t)^{-1} (W_a^t \theta_a + S_a^t) - \theta_a | \\ \nonumber
    &= \sqrt{V_a^t} |(V_a^t)^{-1} S_a^t + \big( (V_a^t)^{-1} W_a^t - 1 \big) \theta_a | \\ \nonumber
    &= \sqrt{V_a^t} |(V_a^t)^{-1} S_a^t + \big( (V_a^t)^{-1} (V_a^t - \lambda_a) - 1 \big) \theta_a | \\ \label{Eqn: Bounding Weighted theta error}
    &= \sqrt{V_a^t} |(V_a^t)^{-1} S_a^t - \lambda_a (V_a^t)^{-1} \theta_a| \\ \nonumber
    &= (V_a^t)^{-1/2} |S_a^t| + \sqrt{\lambda_a} \theta_a.
\end{align}
To bound $(V_a^t)^{-1/2} |S_a^t|$, define $M_a^t(z) := \exp\left( z S_a^t - \frac{1}{2} V_a^t z^2 \right)$ for each $z \in \R$. Then, for any fixed $z \in \R$:
{\small
\begin{align*}
    &\hspace{5mm} \E[M_a^t(z)|\mathcal{F}_a^t] \\
    &= M_a^{t-1}(z) \cdot \E\left[ \exp\left( \arcLoadMod_a^t \cdot \sum_{k=1}^{\lfloor \arcLoadMod_a^t \rfloor} \epsilon_{a,k}^t z - \frac{1}{2} \lfloor \arcLoadMod_a^t \rfloor (\arcLoadMod_a^t)^2 z^2 \right) \Bigg| \mathcal{F}_a^t \right] \\
    &= M_a^{t-1}(z) \cdot \prod_{k=1}^{\lfloor \arcLoadMod_a^t \rfloor} \E\left[ \exp\left( \arcLoadMod_a^t \cdot  \epsilon_{a,k}^t z - \frac{1}{2} \lfloor \arcLoadMod_a^t \rfloor (\arcLoadMod_a^t)^2 z^2 \right) \Bigg| \mathcal{F}_a^t \right] \\
    &\leq M_a^{t-1}(z).
\end{align*}
}
so $M_a^t(z)$ is a supermartingale adapted to the filtration $\mathcal{F}_a^t := \sigma(\arcLoadMod_a^1, \tilde s_a^1)$. Thus, so is $\tilde M_a^t := \E_{z \sim \mathcal{N}(0,1)}[M_a^t(z)]$. It thus follows from Lattimore and Szepesvari \cite{LattimoreSzepesvari2020BanditAlgorithms}, Theorem 20.4, that:
\begin{align} \label{Eqn: Bounding Weighted S}
    (V_a^t)^{-1/2} |S_a^t| &\leq \sqrt{2 \ln t + \ln\left( \frac{V_a^t}{\lambda_a} \right)}.
\end{align}
The proof now follows from \eqref{Eqn: Bounding Weighted theta error} and \eqref{Eqn: Bounding Weighted S}.
\end{proof}

\subsection{Proof of Lemma \ref{Lemma: Existence of Node for Estimating beta}}
\label{subsec: A1, Proof of Lemma: Existence of Node for Estimating beta}


\begin{proof}(\textbf{Proof of Lemma \ref{Lemma: Existence of Node for Estimating beta}})
By assumption, the graph $\graphMod$ contains more than one route from the origin $o$ to the destination $d$. Thus, there exists some $a \in \arcsMod$ such that $|\arcsMod_{i_a}^+| \geq 2$, so the quantity:
\begin{align*}
    m^\star := \min \{m_a: a \in \arcsMod, |\arcsMod_{i_a}^+| \geq 2 \}
\end{align*}
is well-defined. Now, fix any $a \in \arcsMod$ such that $m_a = m^\star$, and $|\arcsMod_{i_a}^+| \geq 2$. It suffices to show that, for each $j \in \arcsMod_{i_a}^+$, there exists only one route connecting $j$ to the destination $d$. Suppose by contradiction that there exists some $j' \in \arcsMod_{i_a}^+$ such that at least two distinct routes connect $j'$ to $d$. Let $\bar j \in \nodesMod \backslash \{d\}$ denote any node at which these routes diverge. Then for any $\bar a \in A_{\bar j}^\star$, we have $|\arcsMod_{i_{\bar a}}^+| = |\arcsMod_{\bar j}^+| \geq 2$, and:
\begin{align*}
    m_{\bar a} < m_a = m^\star,
\end{align*}
a contradiction to the definition of $m^\star$. This concludes the proof.
\end{proof}

\subsection{Proof of Lemma \ref{Lemma: Beta Estimator}}
\label{subsec: A1, Proof of Lemma: Beta Estimator}

\begin{proof}(\textbf{Proof of Lemma \ref{Lemma: Beta Estimator}})
Fix $t \in [T]$. Define $\kappa_{a^\star}^t \in \R$ by:
\begin{align*}
    \kappa_{a^\star}^t &:= \frac{\exp(-\beta^\star \cdot z_{a^\star}(\theta^\star, \arcLoadMod^t, \toll^t))}{\sum_{a' \in \arcsMod_{i^\star}^+} \exp(-\beta^\star \cdot z_{a'}(\theta^\star, \arcLoadMod^t, \toll^t))} \\
    &= \frac{\arcLoadMod_{a^\star}^t}{\sum_{a' \in \arcsMod_{i^\star}^+} \arcLoadMod_{a'}^t},
\end{align*}
and let $f^t, g^t: \R \times \R^{|\arcsMod| \times \R^{|\arcsMod|}} \ra \R$ be given as follows:
\begin{align*}
    &\hspace{5mm} f^t(\beta, \theta^+, \theta^-) \\
    &:= \frac{\exp(-\beta^t \cdot z_{a^\star}(\theta^{t,-}, \arcLoadMod^t, \toll^t))}{\sum_{a' \in \arcsMod_{i^\star}^+} \exp(-\beta^t \cdot z_{a'}(\theta^{t,+}, \arcLoadMod^t, \toll^t))}, \\
    &\hspace{5mm} g^t(\beta, \theta^+, \theta^-) \\
    &:= \ln f^t(\beta, \theta^+, \theta^-) - \ln \kappa_a^t \\
    &= - \beta \cdot z_{a^\star}(\theta^-, \arcLoadMod^t, \toll^t) \\
    &\hspace{1cm} - \ln\left( \sum_{a' \in \arcsMod_{i^\star}^+} \exp \left( -\beta \cdot z_{a'}(\theta^+, \arcLoadMod^t, \toll^t) \right) \right) - \ln \kappa_{a^\star}^t
\end{align*}
Note that $g^t(\beta^t, \theta^+, \theta^-) = 0$ holds if and only if $f^t(\beta^t, \theta^+, \theta^-) = \kappa_a^t$. If one takes $\theta^+ = \theta^{t,+}$ and $\theta^- = \theta^{t,+}$ this becomes a restatement of \eqref{Eqn: Fixed Point Equation}. We note that $z_a(\theta, \arcLoadMod, \toll)$ is continuously differentiable for each $a \in \arcsMod$, $\theta \in \R^{|\arcsMod|}$, $\arcLoadMod \in \arcsLoadConstraintSet$, and $\toll \in \R^{|\arcsMod|}$, and the log-sum-exp function is continuously differentiable in the entropy parameter $\beta$. Thus, $f^t$ and $g^t$ are likewise continuously differentiable at each $\beta > 0$ and each $\theta^+, \theta^- \in \R^{|\arcsMod|}$.

The remainder of the proof proceeds in two parts. We first prove that, given any fixed values $\theta_a^+ \geq \theta_a^\star$, $\theta_a^- \leq \theta_a^\star$ for each $a \in \arcsMod$, there exists a unique fixed point solution $\beta$ to the function $g^t(\beta^t, \theta^+, \theta^-) = 0$. In particular, given $\theta_a^{t,+} \geq \theta_a^\star$, $\theta_a^{t,-} \leq \theta_a^\star$ for each $a \in \arcsMod$, there exists a unique entropy parameter estimate $\beta^t > 0$ that solves $g^t(\beta^t, \theta^{t,+}, \theta^{t,-}) = 0$, i.e., that satisfies \eqref{Eqn: Fixed Point Equation}, and $\beta^\star$ is the unique entropy parameter value that satisfies $g^t(\beta^\star, \theta^\star, \theta^\star) = 0$. We then bound the gap between $\beta^\star$ and $\beta$ by bounding the difference between $\theta^{t,+}$ and $\theta^\star$, and between $\theta^{t,-}$ and $\theta^\star$.

\begin{enumerate}
    \item Claim---Given any fixed $\theta_a^+ \geq \theta_a^\star$, $\theta_a^- \leq \theta_a^\star$ for each $a \in \arcsMod$, there exists a unique fixed point solution $\beta$ to the function $g^t(\beta^t, \theta^+, \theta^-) = 0$:

    $\hspace{1cm}$ \textit{Proof}: To show that, for any $\theta^{t,+}, \theta^{t,-} \in \R^{|\arcsMod|}$, the fixed-point equation $g^t(\beta^\star, \theta^{t,+}, \theta^{t,-}) = 0$, has a unique solution (or equivalently that $f^t(\beta, \theta^+, \theta^-) = \kappa_a^t$ has a unique solution), we first note that:
    \begin{align*}
        \frac{1}{|\arcsMod_{i^\star}^+|} &\leq \kappa_{a^\star}^t \\
        &= \frac{\exp(-\beta^\star \cdot z_{a^\star}(\theta^\star, \arcLoadMod^t, \toll^t))}{\sum_{a' \in \arcsMod_{i^\star}^+} \exp(-\beta^\star \cdot z_{a'}(\theta^\star, \arcLoadMod^t, \toll^t))} \\
        &< 1.
    \end{align*}
    and that $f^t(0, \theta^+, \theta^-) = 1/|\arcsMod_{i^\star}^+|$. Below, we establish that $\lim_{\beta \ra \infty} f^t(\beta, \theta^+, \theta^-) = 1/|\arcsMod_{i^\star}^+|$, by lower bounding $\frac{\partial g^t}{\partial \beta}$. The existence and uniqueness of a solution $\beta$ to the fixed-point equation $f^t(\beta, \theta^+, \theta^-) = \kappa_{a^\star}^t$ then follows from the Intermediate Value Theorem.
    
    $\hspace{5mm}$ To compute derivatives of $g^t$, we observe that, since $i_{a^\star} = i^\star$ satisfies the conditions of Lemma \ref{Lemma: Existence of Node for Estimating beta}, for each $a' \in \arcsMod_{i^\star}^+$, there exists exactly one route that connects $j_{a'}$ and $d$. As a result, $z_{a'}(\theta^+, \arcLoadMod^t, \toll^t)$ equals the sum of latencies on $a'$ and on arcs comprising that route, and therefore does not depend on the entropy parameter $\beta$. Thus:
    \begin{align*}
        &\hspace{5mm} \frac{\partial g^t}{\partial \beta}(\beta, \theta^+, \theta^-) \\
        &= - z_{a^\star}(\theta^-, \arcLoadMod^t, \toll^t) \\
        &\hspace{1cm} + \frac{\sum\limits_{a' \in \arcsMod_{i^\star}^+} e^{- \beta \cdot z_{a'}(\theta^{t,-}, \arcLoadMod^t, \toll^t)} \cdot z_{a'}(\theta^{t,-}, \arcLoadMod^t, \toll^t)}{\sum\limits_{a' \in \arcsMod_{i^\star}^+} e^{- \beta \cdot z_{a'}(\theta^{t,-}, \arcLoadMod^t, \toll^t)}} \\
        &= - z_{a^\star}(\theta^-, \arcLoadMod^t, \toll^t) \\
        &\hspace{1cm} + \sum_{\bar a \in \arcsMod_{i^\star}^+} \frac{ e^{- \beta \cdot z_{\bar a}(\theta^{t,-}, \arcLoadMod^t, \toll^t)}}{\sum\limits_{a' \in \arcsMod_{i^\star}^+} e^{- \beta \cdot z_{a'}(\theta^{t,-}, \arcLoadMod^t, \toll^t)}} \\
        &\hspace{2cm} \cdot z_{\bar a}(\theta^{t,-}, \arcLoadMod^t, \toll^t) \\
        &= \sum_{\bar a \in \arcsMod_{i^\star}^+} \frac{ e^{- \beta \cdot z_{\bar a}(\theta^{t,-}, \arcLoadMod^t, \toll^t)}}{\sum\limits_{a' \in \arcsMod_{i^\star}^+} e^{- \beta \cdot z_{a'}(\theta^{t,-}, \arcLoadMod^t, \toll^t)}} \\
        &\hspace{1.5cm} \cdot \Big[ z_{\bar a}(\theta^{t,-}, \arcLoadMod^t, \toll^t) - z_{a^\star}(\theta^{t,-}, \arcLoadMod^t, \toll^t) \Big] \\
        &= \sum_{\bar a \in \arcsMod_{i^\star}^+} \frac{\arcLoadMod_{\bar a}^t}{\sum\limits_{a' \in \arcsMod_{i^\star}^+} \arcLoadMod_{a'}^t} \\
        &\hspace{1cm} \cdot \Big[ z_{\bar a}(\theta^{t,-}, \arcLoadMod^t, \toll^t) - z_{a^\star}(\theta^{t,-}, \arcLoadMod^t, \toll^t) \Big].
    \end{align*}
    The flow continuity equations imply that $\sum_{a' \in \arcsMod_{i^\star}^+} \arcLoadMod_{a'}^t \leq g_o$; together with the assumption that $\arcLoadMod_a^t \geq 1$ for each $a \in \arcsMod$, we have:
    \begin{align*}
        \frac{\arcLoadMod_{\bar a}^t}{\sum_{a' \in \arcsMod_{i^\star}^+} \arcLoadMod_{a'}^t} \geq \frac{1}{g_o}.
    \end{align*}
    Combining this with the definition of $\Delta_z$, we obtain:
    \begin{align} \label{Eqn: dg d beta, lower bound}
        \frac{\partial g^t}{\partial \beta}(\beta, \theta^+, \theta^-) \geq \frac{\Delta_z}{g_o}.
    \end{align}
    Thus, $g^t(\beta, \theta^+, \theta^-)$ increases to $+\infty$ as $\beta \ra \infty$, and therefore so does $f^t$.

    $\hspace{1cm}$ To reiterate for emphasis, this claim establishes the unique existence of a entropy parameter estimate $\beta^t > 0$ that satisfies $g^t(\cdot, \theta^{t,+}, \theta^{t,-}) = 0$, or equivalently, \eqref{Eqn: Fixed Point Equation}. This claim also establishes that $\beta = \beta^\star$ is the unique solution to $g^t(\cdot, \theta^\star, \theta^\star) = 0$.

    \item Claim---We have:
    \begin{align*}
        |\beta^t - \beta^\star| = \frac{\beta^\star g_o}{\Delta_z} \cdot \sum_{a \in \arcsMod(i^\star)} (\theta_a^{t,+} - \theta_a^{t,-}) \arcLoadMod_a^t.
    \end{align*}
     
    $\hspace{1cm}$ \textit{Proof}: For convenience, we denote $\theta^{\pm} := (\theta^+, \theta^-) \in \R^{2|\arcsMod|}$. For any $\theta^{\pm} \in \R^{2|\arcsMod|}$ such that $\theta_a^+ > \theta_a^\star$ and $\theta_a^- < \theta_a^\star$ for each $a \in \arcsMod$, let $\beta = \hat \beta(\theta^+, \theta^-)$ denote the unique solution to $g^t(\beta, \theta^+, \theta^-)$. Note that for any fixed $\arcLoadMod \in \arcsLoadConstraintSet$ and $\toll \in \R^{|\arcsMod|}$, since $z_{a^\star}(\theta, \arcLoadMod, \toll)$ is component-wise increasing in $\theta$, we have $f^t(0, \theta^+, \theta^-) \leq \kappa_a^t \leq f^t(\beta^\star, \theta^+, \theta^-)$. It thus follows from the Intermediate Value Theorem that $\hat \beta(\theta^+, \theta^-) \in [0, \beta^\star]$.
    
    $\hspace{1cm}$ By \eqref{Eqn: dg d beta, lower bound}, we have $\frac{\partial g^t}{\partial \beta}(\beta, \theta^+, \theta^-) \ne 0$ at each $\beta > 0$. This allows us to apply the Implicit Function Theorem, which yields that $\hat \beta$ is continuously differentiable in $\theta^{\pm}$, with:
    \begin{align*}
        \frac{\partial \hat \beta}{\partial \theta^{\pm}}(\theta^{\pm}) = \Bigg[ \frac{\partial g}{\partial \beta}(\beta, \theta^+, \theta^-) \Bigg]^{-1} \Bigg[ \frac{\partial g}{\partial \theta^{\pm}}(\beta, \theta^+, \theta^-) \Bigg]
    \end{align*}
    Now, define $u_+ := \theta^{t,+} - \theta^\star$ and $u_- := \theta^{t,-} - \theta^\star$. We then have:
    \begin{align*}
        &\hspace{5mm} |\beta^t - \beta^\star| \\
        &= \left|\int_0^1 \frac{\partial \hat \beta}{\partial \theta^{\pm}}(\theta^+ + \sigma u_+, \theta^- + \sigma u_-)^\top  \hspace{0.5mm} dt \right| \\
        &= \Bigg|\int_0^1 \Bigg[ \frac{\partial g}{\partial \beta}(\beta, \theta^\star - \sigma u_+, \theta^\star + \sigma u_-) \Bigg]^{-1} \\
        &\hspace{1cm} \cdot \frac{\partial g}{\partial \theta^{\pm}}(\beta, \theta^+ + \sigma u_-, \theta^- + \sigma u_+) \\
        &\hspace{2cm} \cdot (\theta^+ - \theta^\star, \theta^- - \theta^\star) \hspace{0.5mm} dt \Bigg| \\
        &\leq \frac{g_o}{\Delta_z} \cdot \int_0^1 \Bigg| \frac{\partial g}{\partial \theta^{\pm}}(\beta, \theta^+ + \sigma u_-, \theta^- + \sigma u_+) \\
        &\hspace{2.5cm} \cdot (\theta^+ - \theta^\star, \theta^- - \theta^\star) \Bigg| \hspace{0.5mm} dt \\
        &= \frac{g_o}{\Delta_z} \cdot \int_0^1 \Bigg| \sum_{a \in \arcsMod} \frac{\partial g}{\partial \theta_a^+}(\beta, \theta^+ + \sigma u_-, \theta^- + \sigma u_+) \\
        &\hspace{3cm} \cdot (\theta_a^{t,+} - \theta^\star) \\
        &\hspace{2cm} + \sum_{a \in \arcsMod} \frac{\partial g}{\partial \theta_a^-}(\beta, \theta^+ + \sigma u_-, \theta^- + \sigma u_+) \\
        &\hspace{3cm} \cdot (\theta_a^{t,-} - \theta^\star) \Bigg| \hspace{0.5mm} dt,
    \end{align*}
    where the inequality follows from \eqref{Eqn: dg d beta, lower bound}. Next, let $A(i^\star)$ denote the set of all arcs along routes from the node $i^\star$ to the destination node $d$. Now, observe that, for any $a \in \arcsMod$, $\beta > 0$ and $\theta^+, \theta^- \in \R^{|\arcsMod|}$:
    {\small
     \begin{align*}
        \frac{\partial g}{\partial \theta_a^+}(\beta, \theta^+, \theta^-) &= - \beta \arcLoadMod_a^t \cdot \textbf{1}\{a \in A(i^\star) \}, \\
        \frac{\partial g}{\partial \theta_a^-}(\beta, \theta^+, \theta^-) &= \frac{ \exp(- \beta \cdot z_{\bar a}(\theta^{t,-}, \arcLoadMod^t, \toll^t))}{\sum_{a' \in \arcsMod_{i^\star}^+} \exp(- \beta \cdot z_{a'}(\theta^{t,-}, \arcLoadMod^t, \toll^t))} \\
        &\hspace{1cm} \cdot \beta \arcLoadMod_a^t \cdot \textbf{1}\{a \in A(i^\star) \}.
    \end{align*}
    }
    Substituting into the above upper bound for $|\beta^t - \beta^\star|$, we obtain:
    \begin{align*}
        &\hspace{5mm} |\beta^t - \beta^\star| \\
        &\leq \frac{g_o}{\Delta_z} \cdot \int_0^1 \sum_{a \in \arcsMod(i^\star)} \Bigg| \frac{\partial g}{\partial \theta_a^+}(\beta, \theta^+ + \sigma u_-, \theta^- + \sigma u_+) \Bigg| \\
        &\hspace{3cm} \cdot (\theta_a^{t,+} - \theta^\star) \\
        &\hspace{1.5cm} + \sum_{a \in \arcsMod(i^\star)} \Bigg| \frac{\partial g}{\partial \theta_a^-}(\beta, \theta^+ + \sigma u_-, \theta^- + \sigma u_+) \Bigg| \\
        &\hspace{3cm} \cdot (\theta^\star - \theta_a^{t,-}) \hspace{0.5mm} dt \\
        &\leq \frac{\beta^\star g_o}{\Delta_z} \cdot \sum_{a \in \arcsMod(i^\star)} (\theta_a^{t,+} - \theta_a^{t,-}) \arcLoadMod_a^t,
    \end{align*}
    as desired.    
\end{enumerate}

\end{proof}

\section{Proofs for Section \ref{sec: Regret Analysis}}
\label{sec: A2, Regret Analysis}

\textit{Notation}: Throughout the appendix, the notation $x \lesssim y$ denotes that there exists some constant $K(\lambda, \Delta_z, c_\theta, C_\theta, c_\beta, \beta^\star)$, such that $x \leq Ky$.

\subsection{Preliminary Lemmas}
\label{subsec: A2, Preliminary Lemmas}

This subsection presents preliminary lemmas that will facilitate the proof of Theorem \ref{Thm: Regret Analysis}. We begin with a result derived from the Fundamental Theorem of Calculus. 

\begin{lemma} \label{Lemma: Fundamental Theorem of Calculus, Applied}
If $f: \R^n \ra \R^m$ is continuously differentiable, then, for each $x_1, x_2 \in \R^n$:
\begin{align*}
    &\hspace{5mm} \Vert f(x_2) - f(x_1) \Vert_2 \\
    &\leq \max_{t \in [0, 1]} \left\Vert  \frac{\partial f}{\partial x}\big(x_1 + t(x_2 - x_1) \big) \right\Vert_2 \cdot \Vert x_2 - x_1 \Vert_2.
\end{align*}
\end{lemma}

\begin{proof}
Fix $x_1, x_2 \in \R^n$. For each $i \in [m]$, let $f_i: \R^n \ra \R$ denote the $i$-th component of the map $f$. Define $g_i: \R \ra \R$ by:
\begin{align*}
    g_i(t) := f\big(x_1 + t(x_2 - x_1) \big).
\end{align*}
Then, for each $x_1, x_2 \in \R^n$ and each $i \in [m]$:
\begin{align*}
    &\hspace{5mm} f_i(x_2) - f_i(x_1) \\
    &= g(1) - g(0) \\
    &= \int_0^1 \frac{dg_i}{dt}(t) \hspace{0.5mm} dt \\
    &= \int_0^1 \frac{\partial f_i}{\partial x} \big(x_1 + t(x_2 - x_1) \big) \hspace{0.5mm} dt \cdot (x_2 - x_1).
\end{align*}
Concatenating the above equality across $i \in [m]$, we obtain:
\begin{align*}
    f(x_2) - f(x_1) &= \int_0^1 \frac{\partial f}{\partial x} \big(x_1 + t(x_2 - x_1) \big) \hspace{0.5mm} dt \cdot (x_2 - x_1).
\end{align*}
Finally, we apply the Cauchy-Schwarz inequality to obtain:
\begin{align*}
    &\hspace{5mm} \Vert f(x_2) - f(x_1) \Vert_2 \\
    &\leq \int_0^1 \left\Vert \frac{\partial f}{\partial x} \big(x_1 + t(x_2 - x_1) \big) \right\Vert_2 \hspace{0.5mm} dt \cdot \Vert x_2 - x_1 \Vert_2 \\
    &\leq \max_{t \in [0, 1]} \left\Vert  \frac{\partial f}{\partial x}\big(x_1 + t(x_2 - x_1) \big) \right\Vert_2 \cdot \Vert x_2 - x_1 \Vert_2,
\end{align*}
as desired.
\end{proof}

Below, we establish a collection of upper bounds that will be used repeatedly throughout the remainder of the proofs (Lemmas \ref{Lemma: gamma, upper bound} and \ref{Lemma: Sum of logs, upper bound}).

\begin{lemma} \label{Lemma: gamma, upper bound}
For any $a \in \arcsMod$ and $t \in [T]$:
\begin{align*}
    \gamma_a^t \lesssim \sqrt{\ln(T g_o)}.
\end{align*}
\end{lemma}

\begin{proof}
Recall the definition of $\gamma_a^t$ in \eqref{Eqn: gamma}. After taking $\lambda_a = 1$, we have, for any $t \geq 2$:
\begin{align*}
    \gamma_a^t &= \sqrt{\lambda_a} C_\theta + \sqrt{2 \ln T + 2 \ln \left( \frac{V_a^{t-1}}{\lambda_a} \right)} \\
    &= C_\theta + \sqrt{2 \ln T + 2 \ln\left( 1 + \sum_{t=1}^{t-1} \lfloor \arcLoadMod_a^t \rfloor (\arcLoadMod_a^t)^2 \right)} \\
    &\leq C_\theta + \sqrt{2 \ln T + 2 \ln\big(1 + (t-1) g_o^3 \big)} \\
    &\lesssim \sqrt{\ln(T g_o)}.
\end{align*}
This result can be straightforwardly extended to the $t = 1$ case by ensuring that the constant encapsulated in the \say{$\lesssim$} is selected to be large enough.
\end{proof}

\begin{lemma} \label{Lemma: Sum of logs, upper bound}
For any $a \in \arcsMod$:
\begin{align*}
    \sum_{t=1}^T \min\Bigg\{ 1, \frac{\lfloor \arcLoadMod_a^t \rfloor (\arcLoadMod_a^t)^2}{V_a^{t-1}} \Bigg\} \lesssim \ln(T g_o).
\end{align*}
\end{lemma}

\begin{proof}
First, observe that $\min\{1, x\} \leq \frac{1}{\ln 2} \cdot \ln(1+x)$ for each $x \geq 0$. Thus:
\begin{align*}
    &\sum_{t=1}^T \min\Bigg\{ 1, \frac{\lfloor \arcLoadMod_a^t \rfloor (\arcLoadMod_a^t)^2}{V_a^{t-1}} \Bigg\} \\
    \leq \hspace{0.5mm} &\frac{1}{ \ln 2} \cdot \sum_{t=1}^T \ln\Bigg( 1 + \frac{\lfloor \arcLoadMod_a^t \rfloor (\arcLoadMod_a^t)^2}{V_a^{t-1}} \Bigg) \\
    = \hspace{0.5mm} &\frac{1}{ \ln 2} \cdot \sum_{t=1}^T \ln\Bigg( \frac{V_a^{t-1} + \lfloor \arcLoadMod_a^t \rfloor (\arcLoadMod_a^t)^2}{V_a^{t-1}} \Bigg)
    \\
    = \hspace{0.5mm} &\frac{1}{ \ln 2} \cdot \sum_{t=1}^T \ln\Bigg( \frac{V_a^{t-1} + \lfloor \arcLoadMod_a^t \rfloor (\arcLoadMod_a^t)^2}{V_a^{t-1}} \Bigg)
    \\
    \leq \hspace{0.5mm} &\frac{1}{ \ln 2} \cdot \sum_{t=1}^T \ln\Bigg( \frac{V_a^t}{V_a^{t-1}} \Bigg) \\
    = \hspace{0.5mm} &\frac{1}{ \ln 2} \cdot \ln V_a^T \\
    \leq \hspace{0.5mm} &\frac{1}{\ln 2} \cdot \ln\left(1 + T g_o^3 \right) \\
    \lesssim \hspace{0.5mm} &\ln(T g_o),
\end{align*}
as desired.
\end{proof}

Next, we bound the weighted sums of the magnitudes of the latency function parameter errors $ \theta^{t,-} - \theta^\star$ and entropy parameter $\beta^t - \beta^\star$ across iterations $t \in [T]$. 
First, we require the following lemma.

\begin{lemma} \label{Lemma: theta error sum, upper bounds}
Under the good event $E$, for any $p > 0$:
\begin{align} \label{Eqn: theta error sum, upper bound, star}
    \sum_{t=1}^T \sum_{a \in \arcsMod} |\theta_a^{t,-} - \theta_a^\star| (\arcLoadMod_a^t)^p &\lesssim g_o^p |\arcsMod| \sqrt{T} \ln(T g_o). 
\end{align}
\end{lemma}

\begin{proof}
The desired result follows by taking $p = 2$ in Lemma \ref{Lemma: theta error sum, upper bounds}.
\end{proof}

\begin{lemma} \label{Lemma: beta error sum, upper bound}
Recall that $B$ denotes the number of arcs along routes from $i^\star$ to $d$, which are used to construct an estimate of $\beta^\star$ at each iteration $t$. Under the good event $E$:
\begin{align*}
    \sum_{t=1}^T |\beta^t - \beta^\star| \lesssim g_o B \sqrt{T} \ln(T g_o).
\end{align*}
\end{lemma}

\begin{proof}
Let $\arcsMod(i^\star)$ denote the set of all arcs on routes from $i^\star$ to $d$. By Lemma \ref{Lemma: Beta Estimator}, under the good event $E$, we have $\beta^t \in [c_\beta, \beta^\star]$, so $|\beta^t - \beta^\star| \leq \beta^\star - c_\beta$. Moreover, from \eqref{Eqn: Upper Bound for beta estimate, using theta estimate}, we have:
\begin{align*}
    |\beta^t - \beta^\star| \lesssim g_o \cdot \sum_{a \in \arcsMod(i^\star)} (\theta_a^{t,+} - \theta_a^{t,-}) \arcLoadMod_a^t
\end{align*}
We then have:
\begin{align*}
    &\sum_{t=1}^T |\beta^t - \beta^\star| \\
    \lesssim \hspace{0.5mm} &g_o \cdot \sum_{t=1}^T \left| \min\left\{ \beta^\star - c_\beta, \sum_{a \in \arcsMod(i^\star)} (\theta_a^{t,+} - \theta_a^{t,-}) \arcLoadMod_a^t \right\} \right| \\
    \lesssim \hspace{0.5mm} &g_o 
    \cdot \sum_{t=1}^T \min\left\{ 1, \sum_{a \in \arcsMod(i^\star)}(\theta_a^{t,+} - \theta_a^{t,-}) \arcLoadMod_a^t \right\}.
\end{align*}
Take $\tilde a \in \max_{a \in \arcsMod(i^\star)} \big\{ \sum_{t=1}^T (\theta_a^{t,+} - \theta_a^{t,-}) \arcLoadMod_a^t \big\}$. Then:
\begin{align*}
    &\sum_{t=1}^T |\beta^t - \beta^\star| \\
    \lesssim \hspace{0.5mm} &g_o \cdot \sum_{t=1}^T \min\left\{ 1, B \cdot \frac{2 \gamma_{\tilde a}^t}{\sqrt{V_{\tilde a}^{t-1}}} \arcLoadMod_a^t \right\} \\
    \leq \hspace{0.5mm} &4g_o B \gamma_{\tilde a}^T \cdot \sqrt{T} \cdot \sqrt{\sum_{t=1}^T \min\left\{ 1, \frac{1}{V_{\tilde a}^{t-1}} (\arcLoadMod_a^t)^2 \right\}} \\
    \leq \hspace{0.5mm} &4g_o B \gamma_{\tilde a}^T \cdot \sqrt{T} \cdot \sqrt{\sum_{t=1}^T \min\left\{ 1, \frac{ \lfloor \arcLoadMod_a^t \rfloor (\arcLoadMod_a^t)^2}{V_{\tilde a}^{t-1}} \right\}} \\
    \lesssim \hspace{0.5mm} &g_o B \sqrt{T} \ln(T g_o)
\end{align*}
where we have used the fact that $\lfloor \arcLoadMod_a^t \rfloor \geq 1$.
\end{proof}

\subsection{Upper Bound for \texorpdfstring{$R_1$}{R1}}
\label{subsec: A2, Upper Bound for R1}

\begin{lemma} \label{Lemma: R1, Upper Bound}
Under the good event $E$:
\begin{align} \label{Eqn: R1, Upper Bound}
    R_1 &:= \sum_{t=1}^T \sum_{a \in \arcsMod} (\theta_a^\star - \theta_a^{t,-}) (\arcLoadMod_a^t)^2 \\ \nonumber
    &\lesssim g_o^2 |\arcsMod| \sqrt{T} \ln(T g_o).
\end{align}
\end{lemma}

\begin{proof}
Take $\tilde a \in \arg\max_{a \in \arcsMod} \big\{ \sum_{t=1}^T (\theta_a^\star - \theta_a^{t,-}) (\arcLoadMod_a^t)^2 \big\}$. Then, under the good event $E$:
\begin{align*}
    R_1 &\leq |\arcsMod| \cdot \sum_{t=1}^T (\theta_{\tilde a}^\star - \theta_{\tilde a}^{t,-}) (\arcLoadMod_a^t)^2 \\
    &\leq |\arcsMod| \sqrt{g_o} \cdot \sum_{t=1}^T (\theta_{\tilde a}^\star - \theta_{\tilde a}^{t,-}) (\arcLoadMod_{\tilde a}^t)^{3/2} \\
    &\leq |\arcsMod| \sqrt{g_o} \cdot \sum_{t=1}^T \min\left\{ C_\theta g_o^{3/2}, \frac{2 \gamma_{\tilde a}^t}{\sqrt{V_{\tilde a}^{t-1}}} (\arcLoadMod_{\tilde a}^t)^{3/2} \right\} \\
    &\leq 2\sqrt{2}|\arcsMod| \sqrt{g_o} \\
    &\hspace{1cm} \cdot \sum_{t=1}^T \min\left\{ C_\theta g_o^{3/2}, \frac{\gamma_{\tilde a}^t}{\sqrt{V_{\tilde a}^{t-1}}}  \cdot \sqrt{\lfloor \arcLoadMod_{\tilde a}^t \rfloor} \cdot \arcLoadMod_{\tilde a}^t \right\}.
\end{align*}
where in the final inequality, we have used the fact that, since $\arcLoadMod_a^t \geq 1$ by assumption, we have $\arcLoadMod_a^t \leq 2 \lfloor \arcLoadMod_a^t \rfloor$. Thus, the Cauchy-Schwarz inequality gives:
\begin{align*}
    R_1 &\leq 2\sqrt{2} C_\theta |\arcsMod| g_o^2 \gamma_{\tilde a}^T \\
    &\hspace{1cm} \cdot \sum_{t=1}^T \min\left\{ 1, \frac{1}{\sqrt{V_{\tilde a}^{t-1}}}  \cdot \sqrt{\lfloor \arcLoadMod_{\tilde a}^t \rfloor} \cdot \arcLoadMod_{\tilde a}^t \right\} \\
    &\lesssim |\arcsMod| g_o^2 \sqrt{\ln(T g_o)} \cdot \sqrt{T} \cdot \sqrt{\sum_{t=1}^T \min\left\{ 1, \frac{ \lfloor \arcLoadMod_{\tilde a}^t \rfloor (\arcLoadMod_{\tilde a}^t)^2}{V_{\tilde a}^{t-1}} \right\}} \\
    &\lesssim g_o^2 |\arcsMod| \sqrt{T} \ln(T g_o),
\end{align*}
where the final inequality follows from \eqref{Lemma: Sum of logs, upper bound}.

\end{proof}

\subsection{Upper Bound for \texorpdfstring{$R_2$}{R2}}
\label{subsec: A2, Upper Bound for R2}

Recall that in \eqref{Eqn: chi, entropy term}, we defined the entropy term $\chi: \arcsLoadConstraintSet \ra \R$ as follows:
\begin{align*}
    &\chi(\arcLoadMod) \\
    := \hspace{0.5mm} &\sum_{i \in \nodesMod \backslash \{d\}} \left[ \sum_{a \in \arcsMod_i^+} \arcLoadMod_a \ln \arcLoadMod_a - \left( \sum_{a \in \arcsMod_i^+} \arcLoadMod_a \right) \ln \left( \sum_{a \in \arcsMod_i^+} \arcLoadMod_a \right) \right]
\end{align*}

\begin{lemma} \label{Lemma: chi, upper bound}
For any $\arcLoadMod \in \arcsLoadConstraintSet$, we have:
\begin{align*}
    |\chi(\arcLoadMod)| \leq g_o \cdot (|\nodesMod| - 1) \ln\left(\frac{|\arcsMod|}{|\nodesMod| - 1} \right)
\end{align*}
\end{lemma}

\vspace{1mm}
\begin{proof}
First, fix $D > 0$ arbitrarily, and consider the following constrained optimization problem on $\R^d$:
\begin{align*}
    \min_{x \in \R^d} \hspace{1cm} &\sum_{i=1}^d x_i \ln x_i - \left( \sum_{i=1}^d x_i \right) \ln \left( \sum_{i=1}^d x_i \right) \\
    \text{s.t.} \hspace{1cm} &\sum_{i=1}^d x_i = D.
\end{align*}
The Lagrangian of the above problem is given by:
\begin{align*}
    \mathcal{L}(x, \lambda, \mu) &= \sum_{i=1}^d x_i \ln x_i - \left( \sum_{i=1}^d x_i \right) \ln \left( \sum_{i=1}^d x_i \right) \\
    &\hspace{1cm} + \lambda \left( \sum_{i=1}^d x_i - D \right) + \sum_{i=1}^d \mu_i x_i.
\end{align*}
The corresponding KKT conditions are therefore:
\begin{align*}
    0 &= \frac{\partial \mathcal{L}}{\partial x_i} = \ln x_i + 1 - \ln \left( \sum_{j=1}^d x_j \right) - 1 + \lambda + \mu_i \\
    &= \ln \left( \frac{x_i}{\sum_{j=1}^d x_j} \right) + \lambda + \mu_i, \hspace{1cm} \forall \hspace{0.5mm} i \in [d], \\
    0 &= \mu_i x_i, \hspace{1cm} \forall \hspace{0.5mm} i \in [d],
\end{align*}
and $\sum_{i=1}^d x_i = D$. The optimal solution is thus $x^\star = \frac{D}{d}(1, \cdots, 1)$, with corresponding minimum value:
\begin{align*}
    &\hspace{5mm} \sum_{i=1}^d x_i^\star \ln x_i^\star - \left( \sum_{i=1}^d x_i^\star \right) \ln \left( \sum_{i=1}^d x_i^\star \right) \\
    &= d \cdot \frac{D}{d} \ln\left( \frac{D}{d} \right) - D \ln D \\
    &= - D \ln d.
\end{align*}
This implies that:
\begin{align*}
    &\left| \sum_{a \in \arcsMod_i^+} \arcLoadMod_a \ln \arcLoadMod_a - \Bigg( \sum_{a \in \arcsMod_i^+} \arcLoadMod_a \Bigg) \ln \Bigg( \sum_{a \in \arcsMod_i^+} \arcLoadMod_a \Bigg) \right| \\
    \leq \hspace{0.5mm} &\sum_{a \in \arcsMod_i^+} \arcLoadMod_i \cdot \ln |\arcsMod_i^+|.
\end{align*}
Summing over all non-destination nodes, we obtain:
\begin{align*}
    &\left| \sum_{i \in \nodesMod \backslash \{d\}} \left[ \sum_{a \in \arcsMod_i^+} \arcLoadMod_a \ln \arcLoadMod_a - \Bigg( \sum_{a \in \arcsMod_i^+} \arcLoadMod_a \Bigg) \ln \Bigg( \sum_{a \in \arcsMod_i^+} \arcLoadMod_a \Bigg) \right] \right| \\
    \leq \hspace{0.5mm} &\sum_{i \in \nodesMod \backslash \{d\}} \left( \sum_{a \in \arcsMod_i^+} \arcLoadMod_a \right) \ln |\arcsMod_i^+| \\
    \leq \hspace{0.5mm} &g_o \cdot \sum_{i \in \nodesMod \backslash \{d\}} \ln|\arcsMod_i^+| \\
    \leq \hspace{0.5mm} &g_o \cdot |\nodesMod \backslash \{d\}| \ln\left( \prod_{i \in \nodesMod \backslash \{d\}} \ln|\arcsMod_i^+|^{1/|\nodesMod \backslash \{d\}|} \right) \\
    \leq \hspace{0.5mm} &g_o \cdot |\nodesMod \backslash \{d\}| \ln \left( \frac{1}{|\nodesMod \backslash \{d\}|} \sum_{i \in \nodesMod \backslash \{d\}} |\arcsMod_i^+| \right) \\
    = \hspace{0.5mm} &g_o \cdot (|\nodesMod| - 1) \ln \left( \frac{|\arcsMod|}{|\nodesMod| - 1} \right),
\end{align*}
where the final inequality follows from the arithmetic-geometric inequality.
\end{proof}

\begin{lemma} \label{Lemma: R2, Upper Bound}
Under the good event $E$:
\begin{align} \label{Eqn: R2, Upper Bound}
    R_2 &:= \sum_{t=1}^T \left(\frac{1}{\beta^t} - \frac{1}{\beta^\star} \right) \cdot \chi(\arcLoadMod^t) \\ \nonumber
    &\lesssim g_o^2 \cdot B (|\nodesMod| - 1) \ln \left( \frac{|\arcsMod|}{|\nodesMod| - 1} \right) \cdot \sqrt{T} \ln(T g_o).
\end{align}
\end{lemma}

\begin{proof}
From Lemma \ref{Lemma: beta error sum, upper bound}:
\begin{align*}
    \sum_{t=1}^T |\beta^t - \beta^\star| &\lesssim g_o B \sqrt{T} \ln(T g_o).
\end{align*}
This bound, together with the upper bound on $\chi$ provided by Lemma \ref{Lemma: chi, upper bound}, completes the proof.
\end{proof}

\subsection{Upper Bound for \texorpdfstring{$R_3$}{R3}}
\label{subsec: A2, Upper Bound for R3}

\begin{lemma} \label{Lemma: R3, Upper Bound}
Under the good event $E$:
\begin{align} \label{Eqn: R3, Upper Bound}
    R_3 &:= \sum_{t=1}^T \big| \latencyTotal\big( \bar \arcLoadMod^{\theta^\star, \beta^\star}(\toll^t), \theta^{t,-}, \beta^t \big) \\
    &\hspace{1.5cm} - \latencyTotal\big(\bar \arcLoadMod^{\theta^{t,-}, \beta^t}(\toll^t), \theta^{t,-}, \beta^t \big) \big| \\ \nonumber
    &\lesssim g_o^2 \ln^2(g_o) |\arcsMod| \sqrt{T} \ln(T g_o) \cdot \max\left\{|\nodesMod| \ln\left( \frac{|\arcsMod|}{|\nodesMod|} \right), B \right\}.
\end{align}
\end{lemma}

\begin{proof}
Define the map $\tilde \arcLoadMod: \R^{|\arcsMod|} \times \R \times \R^{|\arcsMod|} \ra \R^{|\arcsMod|}$ by $\tilde \arcLoadMod(\theta, \beta, \toll) := \bar \arcLoadMod^{\theta, \beta}(\toll)$. Observe that $L(\cdot, \theta, \beta)$ is continuously differentiable on $\arcsLoadConstraintSet$, for any fixed $\theta \in \R^{|\arcsMod|}$, $\beta > 0$; 
 later, we will establish that $\tilde \arcLoadMod$ is continuously differentiable as well. Then, from the Fundamental Theorem of Calculus to the maps $\latencyTotal$ and $\tilde \arcLoadMod$, we obtain:
\begin{align} \nonumber
    &\latencyTotal\big( \bar \arcLoadMod^{\theta^\star, \beta^\star}(\toll^t), \theta^{t,-}, \beta^t \big) - \latencyTotal\big(\bar \arcLoadMod^{\theta^{t,-}, \beta^t}(\toll^t), \theta^{t,-}, \beta^t \big) \\ \nonumber
    = \hspace{0.5mm} &\big[ \latencyTotal\big( \bar \arcLoadMod^{\theta^\star, \beta^\star}(\toll^t), \theta^{t,-}, \beta^t \big) - \latencyTotal\big(\bar \arcLoadMod^{\theta^{t,-}, \beta^\star}(\toll^t), \theta^{t,-}, \beta^t \big) \big] \\ \nonumber
    &\hspace{5mm} + \big[ \latencyTotal\big( \bar \arcLoadMod^{\theta^{t,-}, \beta^\star}(\toll^t), \theta^{t,-}, \beta^t \big)\\
    &\hspace{2cm} - \latencyTotal\big(\bar \arcLoadMod^{\theta^{t,-}, \beta^t}(\toll^t), \theta^{t,-}, \beta^t \big) \big] \\ \label{Eqn: C difference, using Fundamental Theorem of Calculus}
    = \hspace{0.5mm} &\int_0^1 \frac{\partial \latencyTotal}{\partial \arcLoadMod} \Big( \bar \arcLoadMod^{\theta^{t,-} + u(\theta^\star - \theta^{t,-}), \beta^t}(\toll^t), \theta^{t,-}, \beta^t \Big) \\ \nonumber
    &\hspace{1cm} \cdot \frac{\partial \tilde \arcLoadMod}{\partial \theta} \big(\theta^{t,-} + u(\theta^\star - \theta^{t,-}), \beta^t, \toll^t \big) \hspace{0.5mm} du \\
    &\hspace{1cm} \cdot (\theta^\star - \theta^{t,-}) \\ \nonumber
    &\hspace{5mm} + \int_0^1 \frac{\partial \latencyTotal}{\partial \arcLoadMod} \Big( \bar \arcLoadMod^{\theta^{t,-}, \beta^t + u(\beta^\star - \beta^t)}(\toll^t), \theta^{t,-}, \beta^t \Big) \\ \nonumber
    &\hspace{1.5cm} \cdot \frac{\partial \tilde \arcLoadMod}{\partial \theta} \big(\theta^{t,-}, \beta^t + u(\beta^\star - \beta^t), \toll^t \big) \hspace{0.5mm} du \\
    &\hspace{1cm} \cdot (\beta^\star - \beta^{t,-}).
\end{align}
For convenience, define:
\begin{align*}
    S_{\arcLoadMod, \theta} &:= \Big\{ \bar \arcLoadMod^{\theta^{t,-} + u(\theta^\star - \theta^{t,-}), \beta^t}(\toll^t): u \in [0, 1] \Big\}, \\
    S_{\arcLoadMod, \beta} &:= \Big\{ \bar \arcLoadMod^{\theta^{t,-}, \beta^t + u(\beta^\star - \beta^t)}(\toll^t): u \in [0, 1] \Big\}, \\
    S &:= S_{\arcLoadMod, \theta} \cup S_{\arcLoadMod, \beta}, \\
    S_\theta &:= \Big\{\theta^{t,-} + u (\theta^\star - \theta^{t,-}): u \in [0, 1] \Big\} \\
    S_\beta &:= \Big\{\beta^t + u (\beta^\star - \beta^t): u \in [0, 1] \Big\}.
\end{align*}
Then, by applying the Cauchy-Schwarz inequality to \eqref{Lemma: Fundamental Theorem of Calculus, Applied}, we obtain:
\begin{align} \nonumber
    &\latencyTotal\big( \tilde \arcLoadMod^{\theta^\star, \beta^\star}(\toll^t), \theta^{t,-}, \beta^t \big) - \latencyTotal\big(\tilde \arcLoadMod^{\theta^{t,-}, \beta^t}(\toll^t), \theta^{t,-}, \beta^t \big) \\ \label{Eqn: C difference, bound, terms}
    \leq \hspace{0.5mm} & \max_{\arcLoadMod \in S_\arcLoadMod} \left\Vert \frac{\partial \latencyTotal} {\partial \arcLoadMod} ( \arcLoadMod, \theta^{t,-}, \beta^t ) \right \Vert_2 \\ \nonumber
    &\hspace{1cm} \cdot \Bigg[ \max_{\theta \in S_\theta} \left\Vert \frac{\partial \tilde \arcLoadMod}{\partial \theta} (\theta, \beta^\star, \toll^t) \cdot (\theta^\star - \theta^{t,-}) \right\Vert_2 \\ \nonumber
    &\hspace{2cm} + \max_{\beta \in S_\beta} \left\Vert \frac{\partial \tilde \arcLoadMod}{\partial \beta} (\theta^\star, \beta, \toll^t) \right\Vert_2 \cdot |\beta^\star - \beta^t| \Bigg]
\end{align}
We bound each of the max terms in \eqref{Eqn: C difference, bound, terms} below.

\begin{enumerate}
    \item Bounding $\max_{\arcLoadMod \in S_\arcLoadMod} \left\Vert \frac{\partial \latencyTotal} {\partial \arcLoadMod} ( \arcLoadMod, \theta^{t,-}, \beta^t ) \right \Vert_2$:

    $\hspace{5mm}$ For each $a \in \arcsMod$, and any $\arcLoadMod \in \arcsLoadConstraintSet$, $\theta \in \R^{|\arcsMod|}$, and $\beta > 0$:
    \begin{align*}
        \frac{\partial \latencyTotal}{\partial \arcLoadMod_a}(\arcLoadMod, \theta, \beta) &= 2 \theta_a \arcLoadMod_a + \frac{1}{\beta} \ln \left( \frac{\arcLoadMod_a}{\sum_{a' \in \arcsMod_{i_a}^+} \arcLoadMod_{a'}} \right).
    \end{align*}
    Note that $|\theta_a^{t,-}| \leq C_\theta$ for each $a \in \arcsMod$, and that for any $\arcLoadMod \in \arcsLoadConstraintSet$, we have $\Vert \arcLoadMod \Vert_2 \leq \sum_{a \in \arcsMod} \arcLoadMod_a \leq m(G) g_o$. Moreover, by Lemma \ref{Lemma: chi, upper bound}, and the assumption that $\arcLoadMod_a \geq 1$ for each $a \in \arcsMod$ (note that the set $\{\arcLoadMod \in \R^{|\arcsMod|}: \arcLoadMod_a \geq 1, \forall \hspace{0.5mm} a \in \arcsMod \}$ is convex), we have for each $\arcLoadMod \in \arcsLoadConstraintSet$:
    \begin{align} \nonumber
        &\sum_{a \in \arcsMod} \left| \ln \left( \frac{\arcLoadMod_a}{\sum_{a' \in \arcsMod_{i_a}^+} \arcLoadMod_{a'}} \right) \right| \\ \nonumber
        = \hspace{0.5mm} &-\sum_{a \in \arcsMod} \ln \left( \frac{\arcLoadMod_a}{\sum_{a' \in \arcsMod_{i_a}^+} \arcLoadMod_{a'}} \right) \\ \nonumber
        \leq \hspace{0.5mm} &-\sum_{a \in \arcsMod} \arcLoadMod_a \ln \left( \frac{\arcLoadMod_a}{\sum_{a' \in \arcsMod_{i_a}^+} \arcLoadMod_{a'}} \right) \\ \nonumber
        = \hspace{0.5mm} &|\chi(\arcLoadMod)| \\ \label{Eqn: Log ratio sum, upper bound, 1}
        \leq \hspace{0.5mm} &g_o \cdot (|\nodesMod| - 1)\ln\left( \frac{|\arcsMod|}{|\nodesMod| - 1} \right)
    \end{align}
    Meanwhile:
    \begin{align} \label{Eqn: Log ratio sum, upper bound, 2}
        &\sum_{a \in \arcsMod} \left| \ln \left( \frac{\arcLoadMod_a}{\sum_{a' \in \arcsMod_{i_a}^+} \arcLoadMod_{a'}} \right) \right|^2 \leq \ln^2(g_o) |\arcsMod|
    \end{align}
    Thus, we obtain that, for any $\arcLoadMod \in S_w$:
    {\small
    \begin{align} \nonumber
        &\left\Vert \frac{\partial \latencyTotal}{\partial \arcLoadMod_a}(\arcLoadMod, \theta^{t,-}, \beta^t) \right\Vert_2 \\ \label{Eqn: Bounding R3 summand, 1st term}
        \leq \hspace{0.5mm} &2 C_\theta m(G) g_o \\
        &\hspace{3mm} + \frac{1}{c_\beta} \min\Bigg\{\ln(g_o) \sqrt{|\arcsMod|},  g_o (|\nodesMod| - 1)\ln\left( \frac{|\arcsMod|}{|\nodesMod| - 1} \right) \Bigg\}.
    \end{align}
    }

    \item Bounding $\max_{\theta \in S_\theta} \left\Vert \frac{\partial \tilde \arcLoadMod}{\partial \theta} (\theta, \beta^\star, \toll^t) \cdot (\theta^\star - \theta^{t,-}) \right\Vert_2$:

    \vspace{1mm}
    $\hspace{5mm}$ First, we verify that $\tilde \arcLoadMod$ is indeed continuously differentiable, and compute the Jacobians $\frac{\partial \latencyTotal}{\partial \arcLoadMod}$, $\frac{\partial \tilde \arcLoadMod}{\partial \theta}$, and $\frac{\partial \tilde \arcLoadMod}{\partial \beta}$. This requires the results of \cite{Chiu2023DynamicTollinginArcBasedTAMs}, Lemma 1, which we summarize below. Define $F:\arcsLoadConstraintSet\times \R^{|\arcsMod|} \times \R^{|\arcsMod|} \times \R \times \R^{|\arcsMod|} \ra \R$ as follows---For each:
    {\small
    \begin{align*} 
        &\hspace{5mm} F(\arcLoadMod, \theta, \beta, \toll) \\
        &=\sum_{[a] \in A_O} \int_0^{\arcLoadMod_a} \big[ \theta_a z + \toll_a \big] \hspace{0.5mm} dz \\ 
        &\hspace{1mm} + \frac{1}{\beta} \sum_{i \ne d} \Bigg[ \sum_{a \in A_i^+} \arcLoadMod_a \ln \arcLoadMod_a - \Bigg(\sum_{a \in A_i^+} \arcLoadMod_a \Bigg) \ln \Bigg(\sum_{a \in A_i^+} \arcLoadMod_a \Bigg) \Bigg]
    \end{align*}
    }
    Note that $F(\cdot, \theta, \beta, \toll)$ is strongly convex, with parameter at least $c_\theta$.
    
    Next, observe that $\arcsLoadConstraintSet$ is a compact subset of a strict affine subspace in $\R^{|\arcsMod|}$. Let $d$ be the dimension of the smallest affine subspace containing $\arcsLoadConstraintSet$. Then, there exist $M \in \R^{|A| \times |\nodesMod \backslash \{d\}|}$ with orthonormal columns, and $b \in \R^{|\nodesMod \backslash \{d\}|}$ such that:
    \begin{align*}
        \arcsLoadConstraintSet = \{w \in \R^{|A|}: M^\top w + b = 0, \arcLoadMod_a \geq 0, \hspace{0.5mm} \forall \hspace{0.5mm} a \in A \}.
    \end{align*}
    Let $B \in \R^{|\arcsMod| \times (|\arcsMod| - |\nodesMod \backslash \{d\}|)}$ consist of orthonormal columns orthogonal to the columns of $M$. We then use the theory of constrained optimization to completely characterize $\tilde \arcLoadMod(\theta, \beta, \toll) = \bar \arcLoadMod^{\theta, \beta}(\toll)$. In particular, $\arcLoadMod = \tilde \arcLoadMod(\theta, \beta, \toll)$ if and only if the following implicit equation, characterized by the map $J: \R^{|\arcsMod|} \times \R^{|\arcsMod|} \ra \R^{|\arcsMod|}$ defined below, is satisfied:
    \begin{align*}
        J(\arcLoadMod, \theta, \beta, \toll) &:= \begin{bmatrix}
            M^\top w + b \\
            B^\top \nabla_{\arcLoadMod} F(\arcLoadMod, \theta, \beta, \toll)
        \end{bmatrix} = 0.
    \end{align*}
    Moreover, the proof of \cite{Chiu2023DynamicTollinginArcBasedTAMs}, Lemma 1 establishes that, for any fixed $\theta \in \R^{|\arcsMod|}$, $\beta > 0$, $\toll \in \R^{|\arcsMod|}$:
    \begin{align*}
        \frac{\partial J}{\partial \arcLoadMod}(\theta, \beta, \toll) &= \begin{bmatrix}
            M^\top \\
            B^\top \nabla_{\arcLoadMod}^2 F(\arcLoadMod, \theta, \beta, \toll)
        \end{bmatrix} \in \R^{|A| \times |A|}
    \end{align*}
    is non-singular. By the Implicit Function Theorem, this establishes the continuous differentiability of $\tilde \arcLoadMod$. We can then compute $\frac{\partial \tilde \arcLoadMod}{\partial \theta} \in \R^{|\arcsMod| \times |\arcsMod|}$ at any $(\theta, \beta, \toll) \in \R^{|\arcsMod|} \times \R \times \R^{|\arcsMod|}$ as:
    \begin{align*}
        &\hspace{5mm} \frac{\partial \tilde \arcLoadMod}{\partial \theta}(\theta, \beta, \toll) \\
        &= \Bigg[ \frac{\partial J}{\partial \arcLoadMod}(\theta, \beta, \toll) \Bigg]^{-1} \frac{\partial J}{\partial \theta}(\theta, \beta, \toll) \\
        &= \begin{bmatrix}
            M^\top \\
            B^\top \nabla_{\arcLoadMod}^2 F(\arcLoadMod, \theta, \beta, \toll)
        \end{bmatrix}^{-1}
        \begin{bmatrix}
            0 \\
            B^\top \frac{\partial}{\partial \theta} \nabla_{\arcLoadMod} F(\arcLoadMod, \theta, \beta, \toll)
        \end{bmatrix} \\
        &= B(B^\top \nabla_{\arcLoadMod}^2 F(\arcLoadMod, \theta, \beta, \toll) B)^{-1} B^\top \\
        &\hspace{1cm} \cdot \frac{\partial}{\partial \theta} \nabla_{\arcLoadMod} F(\arcLoadMod, \theta, \beta, \toll),
    \end{align*}
    where we have used the fact that by construction, $\begin{bmatrix}
        M & B
    \end{bmatrix}$ is an orthogonal matrix (see \cite{Chiu2023DynamicTollinginArcBasedTAMs}, Appendix A).  
    
    $\hspace{1cm}$ Now, observe that the $(a, a')$-entry of $\frac{\partial}{\partial \beta} \nabla_\arcLoadMod F(\arcLoadMod, \theta, \beta, \toll) \in \R^{|\arcsMod| \times |\arcsMod|}$ is given by:
    \begin{align*}
        \frac{\partial^2}{\partial \theta_{a'} \partial \arcLoadMod_a} F(\arcLoadMod, \theta, \beta, \toll) &= 2 \arcLoadMod_a \cdot \textbf{1}\{a' = a\}, \hspace{5mm} \forall \hspace{0.5mm} a \in \arcsMod,
    \end{align*}
    Substituting back into \eqref{Eqn: C difference, bound, terms} and applying the Cauchy-Schwarz inequality, we obtain that, for each $\theta \in S_\theta$:
    \begin{align*}
        &\frac{\partial \tilde \arcLoadMod}{\partial \theta} (\theta, \beta^\star, \toll^t) \cdot (\theta^\star - \theta^{t,-}) \\
        = \hspace{0.5mm} &B(B^\top \nabla_{\arcLoadMod}^2 F(\arcLoadMod, \theta, \beta, \toll) B)^{-1} B^\top \\
        &\hspace{5mm} \cdot \big( (\theta^\star - \theta^{t,-}) \arcLoadMod_a^t \big)_{a \in \arcsMod}.
    \end{align*}
    Applying the Cauchy-Schwarz inequality, we obtain:
    \begin{align*}
        &\max_{\theta \in S_\theta} \left\Vert \frac{\partial \tilde \arcLoadMod}{\partial \theta} (\theta, \beta^\star, \toll^t) \cdot (\theta^\star - \theta^{t,-}) \right\Vert_2 \\
        \leq \hspace{0.5mm} &\Vert B(B^\top \nabla_{\arcLoadMod}^2 F(\arcLoadMod, \theta, \beta, \toll) B)^{-1} B^\top \Vert_2 \\
        &\hspace{5mm} \cdot \Vert \big( (\theta^\star - \theta^{t,-}) \arcLoadMod_a^t \big)_{a \in \arcsMod} \Vert_2.
    \end{align*}
    Since the columns of $B$ are orthonormal, we have $\Vert B(B^\top \nabla_{\arcLoadMod}^2 F(\arcLoadMod, \theta, \beta, \toll) B)^{-1} B^\top \Vert_2 \leq \Vert \nabla_w^2 F(\arcLoadMod, \theta, \beta, \toll) \Vert_2 \leq 1/c_\theta$. Moreover, we can upper bound $\Vert \big( (\theta^\star - \theta^{t,-}) \arcLoadMod_a^t \big)_{a \in \arcsMod} \Vert_2 \leq \Vert \big( (\theta^\star - \theta^{t,-}) \arcLoadMod_a^t \big)_{a \in \arcsMod} \Vert_1 = \sum_{a \in \arcsMod} (\theta^\star - \theta^{t,-}) \arcLoadMod_a^t$. We thus obtain:
    \begin{align} \nonumber
        &\max_{\theta \in S_\theta} \left\Vert \frac{\partial \tilde \arcLoadMod}{\partial \theta} (\theta, \beta^\star, \toll^t) \cdot (\theta^\star - \theta^{t,-}) \right\Vert_2 \\ \label{Eqn: Bounding R3 summand, 2nd term}
        \leq \hspace{0.5mm} &\frac{1}{c_\theta} \cdot \sum_{a \in \arcsMod} (\theta^\star - \theta^{t,-}) \arcLoadMod_a^t.
    \end{align}
    
    \item Bounding $\max_{\beta \in S_\beta} \left\Vert \frac{\partial \tilde \arcLoadMod}{\partial \beta} (\theta^\star, \beta, \toll^t) \right\Vert_2 \cdot |\beta^\star - \beta^t|$:

    \vspace{1mm}
    $\hspace{5mm}$ In the same manner that we used to compute $\frac{\partial \tilde \arcLoadMod}{\partial \theta}$ above, we can compute $\frac{\partial \tilde \arcLoadMod}{\partial \beta} \in \R^{|\arcsMod|}$ at any $(\theta, \beta, \toll) \in \R^{|\arcsMod|} \times \R \times \R^{|\arcsMod|}$ as:
    \begin{align*}
        &\hspace{5mm} \frac{\partial \tilde \arcLoadMod}{\partial \beta}(\theta, \beta, \toll) \\
        &= \Bigg[ \frac{\partial J}{\partial \arcLoadMod}(\theta, \beta, \toll) \Bigg]^{-1} \frac{\partial J}{\partial \theta}(\theta, \beta, \toll), \\
        &= \begin{bmatrix}
            M^\top \\
            B^\top \nabla_{\arcLoadMod}^2 F(\arcLoadMod, \theta, \beta, \toll)
        \end{bmatrix}^{-1} 
        \begin{bmatrix}
            0 \\
            B^\top \frac{\partial}{\partial \beta} \nabla_{\arcLoadMod} F(\arcLoadMod, \theta, \beta, \toll)
        \end{bmatrix} \\
        &= B(B^\top \nabla_{\arcLoadMod}^2 F(\arcLoadMod, \theta, \beta, \toll) B)^{-1} B^\top \\
        &\hspace{1cm} \cdot \frac{\partial}{\partial \beta} \nabla_{\arcLoadMod} F(\arcLoadMod, \theta, \beta, \toll),
    \end{align*}
    Now, observe that the $a$-th entry of $\frac{\partial}{\partial \beta} \nabla_\arcLoadMod F(\arcLoadMod, \theta, \beta, \toll) \in \R^{|\arcsMod|}$ is:
    \begin{align*}
        \frac{\partial^2}{\partial \beta \partial \arcLoadMod_a} F(\arcLoadMod, \theta, \beta, \toll) &= - \frac{1}{\beta^2} \ln \left( \frac{\arcLoadMod_a}{\sum_{a' \in \arcsMod_{i_a}^+} \arcLoadMod_{a'}} \right).
    \end{align*}
    Using \eqref{Eqn: Log ratio sum, upper bound, 1} and \eqref{Eqn: Log ratio sum, upper bound, 2}, we obtain:
    \begin{align*}
        &\hspace{5mm} \left\Vert \frac{\partial}{\partial \beta} \nabla_\arcLoadMod F(\arcLoadMod, \theta, \beta, \toll) \right\Vert_2 \\
        &\leq \frac{1}{\beta^2} \cdot \min\left\{ \ln(g_o) \sqrt{|\arcsMod|}, g_o (|\nodesMod| - 1) \ln \left( \frac{|\arcsMod|}{|\nodesMod| - 1} \right) \right\}
    \end{align*}
    Finally, we conclude that:
    \begin{align} \nonumber
        &\max_{\beta \in S_\beta} \left\Vert \frac{\partial \tilde \arcLoadMod}{\partial \beta} (\theta^\star, \beta, \toll^t) \right\Vert_2 \cdot |\beta^\star - \beta^t| \\ \nonumber
        \leq \hspace{0.5mm} &\Vert B(B^\top \nabla_{\arcLoadMod}^2 F(\arcLoadMod, \theta, \beta, \toll) B)^{-1} B^\top \Vert_2 \\ \nonumber
        &\hspace{5mm} \cdot \left\Vert \frac{\partial}{\partial \beta} \nabla_\arcLoadMod F(\arcLoadMod, \theta, \beta, \toll) \right\Vert_2 \cdot |\beta^\star - \beta^t| \\ \label{Eqn: Bounding R3 summand, 3rd term}
        \leq \hspace{0.5mm} &\frac{1}{c_\theta \beta^2} \cdot \min\left\{ \ln(g_o) \sqrt{|\arcsMod|}, g_o (|\nodesMod| - 1) \ln \left( \frac{|\arcsMod|}{|\nodesMod| - 1} \right) \right\} \\ \nonumber
        &\hspace{1cm} \cdot |\beta^\star - \beta^t|.
    \end{align}
\end{enumerate}

Substituting \eqref{Eqn: Bounding R3 summand, 1st term}, \eqref{Eqn: Bounding R3 summand, 2nd term}, \eqref{Eqn: Bounding R3 summand, 3rd term} back into \eqref{Eqn: C difference, bound, terms}, we obtain that:
\begin{align} \nonumber
    R_3 &= \sum_{t=1}^T \big| \latencyTotal\big( \bar \arcLoadMod^{\theta^\star, \beta^\star}(\toll^t), \theta^{t,-}, \beta^t \big) \\ \nonumber
    &\hspace{1cm} - \latencyTotal\big(\bar \arcLoadMod^{\theta^{t,-}, \beta^t}(\toll^t), \theta^{t,-}, \beta^t \big) \big| \\ \nonumber
    &\lesssim \left( m(G) g_o + \frac{1}{c_\beta} \cdot g_o \cdot (|\nodesMod| - 1)\ln\left( \frac{|\arcsMod|}{|\nodesMod| - 1} \right) \right) \\ \nonumber
    &\hspace{5mm} \cdot \Bigg[ \sum_{t=1}^T \sum_{a \in \arcsMod} (\theta^\star - \theta^{t,-}) \arcLoadMod_a^t \\ \label{Eqn: R3, upper bound, intermediate, 1}
    &\hspace{1cm} + \min\left\{ \ln^2(g_o) \cdot |\arcsMod|, g_o |\nodesMod| \ln \left( \frac{|\arcsMod|}{|\nodesMod|} \right) \right\} \\ \nonumber
    &\hspace{1.5cm} \cdot \sum_{t=1}^T |\beta^\star - \beta^t| \Bigg]
\end{align}
Applying Lemmas \ref{Lemma: theta error sum, upper bounds} (with $p = 1$) and \ref{Lemma: beta error sum, upper bound}, we obtain:
\begin{align*}
    R_3 &\lesssim \left( m(G) g_o + \min\Bigg\{\ln(g_o) \sqrt{|\arcsMod|}, g_o |\nodesMod| \ln \left( \frac{|\arcsMod|}{|\nodesMod|} \right) \Bigg\} \right) \\
    &\hspace{1cm} \cdot \Bigg[ g_o |\arcsMod| \sqrt{T} \ln(T g_o) \\
    &\hspace{1.5cm} + \min\Bigg\{\ln(g_o) \sqrt{|\arcsMod|}, g_o|\nodesMod| \ln \left( \frac{|\arcsMod|}{|\nodesMod|} \right) \Bigg\} \\
    &\hspace{2cm} \cdot g_o B \sqrt{T} \ln(T g_o) \Bigg] \\
    &\lesssim g_o^2 m(G) |\arcsMod| \sqrt{T} \ln(T g_o) \\
    &\hspace{1cm} + g_o^2 \ln(g_o) m(G) \sqrt{|\arcsMod|} B \sqrt{T} \ln(T g_o) \\
    &\hspace{1cm} + g_o^2 |\arcsMod| |\nodesMod| \ln\left( \frac{|\arcsMod|}{|\nodesMod|} \right) \sqrt{T} \ln(Tg_o) \\
    &\hspace{1cm} + g_o \ln^2(g_o) |\arcsMod|B \sqrt{T} \ln(Tg_o) \\
    &\lesssim g_o^2 \ln^2(g_o) |\arcsMod| \sqrt{T} \ln(T g_o) \\
    &\hspace{1cm} \cdot \max\left\{|\nodesMod| \ln\left( \frac{|\arcsMod|}{|\nodesMod|} \right), B \right\}.
\end{align*}
Note that we have used the fact that $m(G) \leq |\nodesMod|$.
\end{proof}

\subsection{Upper Bound for \texorpdfstring{$R$}{R}}
\label{subsec: A2, Upper Bound for R}

Below, we combine the results of Lemmas \ref{Lemma: R1, Upper Bound}, \ref{Lemma: R2, Upper Bound}, and \ref{Lemma: R3, Upper Bound} in the above sections to conclude our proof of Theorem \ref{Thm: Regret Analysis}.

\begin{proof}[\textbf{Proof of Theorem \ref{Thm: Regret Analysis}}]
From Lemmas \ref{Lemma: R1, Upper Bound}, \ref{Lemma: R2, Upper Bound}, and \ref{Lemma: R3, Upper Bound}, we have:
\begin{align*}
    R_1 &\lesssim g_o^2 |\arcsMod| \sqrt{T} \ln(T g_o), \\
    R_2 &\lesssim g_o^2 \cdot B |\nodesMod| \ln\left( \frac{|\arcsMod|}{|\nodesMod|} \right) \cdot \sqrt{T} \ln(T g_o), \\
    R_3 &\lesssim g_o^2 \ln^2(g_o) |\arcsMod| \sqrt{T} \ln(T g_o) \cdot \max\left\{|\nodesMod| \ln\left( \frac{|\arcsMod|}{|\nodesMod|} \right), B \right\}.
\end{align*}
Note that $R_1 \lesssim R_3$ and $R_2 \lesssim R_3$. We thus conclude that:
\begin{align*}
    R &= R_1 + R_2 + R_3 \\
    &\lesssim g_o^2 \ln^2(g_o) |\arcsMod| \sqrt{T} \ln(T g_o) \cdot \max\left\{|\nodesMod| \ln\left( \frac{|\arcsMod|}{|\nodesMod|} \right), B \right\}.
\end{align*}
\end{proof}











\end{document}